\documentclass[preprint,12pt]{elsarticle}

\usepackage{subfigure}
\usepackage{amssymb}
\usepackage{xcolor, soul}

\newtheorem{thm}{Theorem}

\usepackage{amssymb}
\usepackage{amsmath}

\begin{document}

\begin{frontmatter}



\title{Intelligent backpropagated neural networks application on Couette-Poiseuille flow of variable viscosity in a composite porous channel filled with an anisotropic porous layer}

\author[label1]{Timir Karmakar}\author[label2]{Amrita Mandal}
\affiliation[label1]{organization={Department of Mathematics, National Institute of Technology Meghalaya},
           city={Sohra},
          postcode={793108},
          state={Meghalaya},
        country={India}}

\affiliation[label2]{organization={Department of Mathematics, Birla Institute of Technology Mesra},
           city={Ranchi},
          postcode={835215},
           state={Jharkhand},
           country={India}}



\begin{abstract}
This study examines Couette-Poiseuille flow of variable viscosity within a channel that is partially filled with a porous medium. To enhance its practical relevance, we assume that the porous medium is anisotropic with permeability varying in all directions, making it a positive semidefinite matrix in the momentum equation. We assume the Navier-Stokes equations govern the flow in the free flow region, while the Brinkman-Forchheimer-extended Darcy's equation governs the flow inside the porous medium. The coupled system contains a nonlinear term from the Brinkman-Forchheimer equation. We propose an approximate solution using an iterative method valid for a wide range of porous media parameter values. For both high and low values of the Darcy number, the asymptotic solutions derived from the regular perturbation method and matched asymptotic expansion show good agreement with the numerical results. However, these methods are not effective in the intermediate range. To address this, we employ the artificial Levenberg-Marquardt method with a back-propagated neural network (ALMM-BNN) paradigm to predict the solution in the intermediate range. While it may not provide the exact solutions, it successfully captures the overall trend and demonstrates good qualitative agreement with the numerical results. This highlights the potential of the ALMM-BNN paradigm as a robust predictive tool in challenging parameter ranges where numerical solutions are either difficult to obtain or computationally expensive. The current model provides valuable insights into the shear stress distribution of arterial blood flow, taking into account the variable viscosity of the blood in the presence of inertial effects. It also offers a framework for creating glycocalyx scaffolding and other microfluidic systems that can mimic the biological glycocalyx.  
\end{abstract}



\begin{keyword}
Brinkman-Forchheimer extended Darcy's equation, Artificial Levenberg Marquardt method with a backpropagated neural network (ALMM-BNN), Regular perturbation, Matched asymptotic expansion



\end{keyword}

\end{frontmatter}



\section{Introduction}\label{sec1}

One classical problem in fluid mechanics is the study of fluid flow in channels that are partially filled with a porous medium. The study of flow through composite porous channels has garnered the attention of researchers due to its practical applications in various fields, including macromolecular transport across an arterial wall, thermal insulation, solid matrix heat exchangers, groundwater transport, and crude oil extraction. The literature in this area can broadly be classified into two categories: one where the flow within the porous medium is governed by Darcy's equation and the other where it is governed by the Brinkman equation. The reliability of using these constitutive models becomes questionable for modeling the flow through porous media characterized by high filtration velocity, where inertial effects play a significant role. This is particularly relevant in areas such as flow near wellbores, fractures, tight-screen cryogenic propellant tanks, karst aquifers, convective flow in carbon nanotubes, film boiling in porous media, and blood flow through arteries \cite{Skrzypacz2017solvability,Arthur2018porous,Karmakar2022analysis,Roure2022modeling}. As the seepage velocity rises, the flow behavior undergoes a gradual transition, resulting in a flow relationship where the pressure gradient is no longer linearly proportional to the seepage velocity. In this situation, it becomes important to account for inertia when describing the flow, as it can no longer be neglected. One possible option to circumvent this deficiency is the inclusion of a non-linear Forchheimer term with the Darcy term, known as Forchheimer's law \cite{forchheimer1901hydraulik,Nield2006convection}. However, Forchheimer's law does not include the Brinkman term, which is crucial for explaining the viscous effect, and also lacks empirical support \cite{Lai2012non}. When both viscous and inertial effects need to be considered, the Brinkman–Forchheimer–extended Darcy model stands out as the most reliable and universally accepted model. Experimental validation of the Brinkman–Forchheimer–extended Darcy model for the case of high flow rates can be found in the study by Kladias \& Prasad \cite{Kladias1991experimental}, Givler \& Altobelli \cite{Givler1994determination}. 
 
A well-known problem in fluid mechanics is the Couette-Poiseuille flow, which occurs between two parallel plates and involves solving the Navier-Stokes equations exactly. In recent years, the investigation of Couette-Poiseuille flow in channels fully or partially filled with a porous medium has garnered significant attention due to its wide range of applications, which include enhanced oil recovery (EOR), filtration and separation processes, polymer and composite material processing, contaminant transport in porous soils, and blood flow through arteries \cite{Nakayama1992non,Yang2006modeling,Alexiou2013plane,Ghosh2020note,Karmakar2021physics}. A key flow scenario in which inertial and boundary effects can play a crucial role is the Couette flow, completely or partially filled with a porous medium. In the Couette flow, the velocity of the moving plate may be substantial, and the viscous forces in the boundary layer near the plate may be significant. It appears in the momentum equation as a quadratic term and becomes more important for large filtration velocities. To accurately describe the flow, it is important to consider non-Darcian effects in the momentum equation. This includes the Forchheimer term to account for the inertial effects and the Brinkman term to account for the viscous effects. The literature on this topic is primarily divided into two categories: one focuses on channels that are completely packed with a porous medium, while the other examines channels that are partially filled with a porous medium. Notable early studies on the former case can be found in the work of Vafai and Tien \cite{Vafai1981boundary}, Vafai and Kim \cite{Vafai1989forced}, Nield et al. \cite{Nield1996forced}, Kuznetsov \cite{Kuznetsov1998analytical}, Nakayama \cite{Nakayama1992non}, Kaloni and Guo \cite{Kaloni1996steady}, and Givler and Altobelli \cite{Givler1994determination}. 

Another interesting area for research is the study of Couette-Poiseuille flows in a composite porous channel that is partially filled with a porous medium. Here, we highlight some of the literature to emphasize advancements in the research as it moves toward increasingly intricate models over time. Vafai and Kim \cite{Vafai1990fluid} offered an exact solution for fluid flow at the interface of a porous medium and a fluid layer, considering inertia and boundary effects. They used the boundary layer approximation to achieve this solution of the momentum equation. Kuznetsov \cite{Kuznetsov1998analytical_composite,Kuznetsov2000fluid} conducted an analytical investigation of the Couette-Poiseuille flow in a composite channel that is partially filled with a porous medium and partially with a clear fluid. They also analyzed the velocity and heat transfer profiles under various interfacial conditions. Their findings indicated the presence of two boundary layers: one near the liquid-porous interface and the other near the bottom of the porous medium. Alazmi and Vafai \cite{Alazmi2001analysis} analyzed various hydrodynamic and thermal interfacial conditions between a porous medium and an adjacent fluid layer. Their findings indicate that the velocity field is more responsive to changes in boundary conditions than the temperature field. Liu et al. \cite{Liu2012poiseuille} studied convective flow and heat transfer in a Couette-Poiseuille flow within an inclined channel that is partially filled with a Brinkman porous medium, considering the effect of viscous dissipation on heat transfer. Hill and Straughan \cite{Hill2008poiseuille} performed a numerical study on the instability of Poiseuille flow in a fluid that lies on top of a porous medium saturated with the same fluid. Alexiou and Kapellos \cite{Alexiou2013plane} presented an analytical model to study Couette-Poiseuille flow over a poroelastic layer. They investigated the role of shear stress distribution in the mechanotransduction process by examining its distribution over the endothelium surface layer. More studies on flow through composite channels partially filled with a porous medium can be found in the research monograph by Nield and Bejan \cite{Nield2006convection}.

Incorporating the inertial (Forchheimer) and viscous (Brinkman) effects into the fully developed momentum transfer equation yields a second-order nonlinear differential equation. The incorporation of the non-linear inertial term in the momentum equation complicates the explicit determination of the velocity field. In the case of a two-layer liquid-porous system, this complexity is further compounded by the fluid flow coupling between the free and porous regions. In addition to increasing the mathematical complexity, the inertial term also modifies the underlying physics of the system. Efforts are being made to obtain a closed-form solution for the non-linear Brinkman-Forchheimer equation in the case of fully developed flow. Vafai and Kim \cite{Vafai1989forced} studied forced convection in a channel that is fully packed with a porous medium, using the Brinkman-Forchheimer extended Darcy's equation to govern the flow. They derived a closed-form solution based on boundary layer approximations. Their study shows that for high-permeability porous media, the thickness of the momentum boundary layer is influenced by both the Darcy number and the inertia parameter.
In contrast, for low-permeability porous media, the thickness depends solely on the Darcy number.  Nield et al. \cite{Nield1996forced} identified several limitations in the study conducted by Vafai and Kim \cite{Vafai1989forced}. They determined that the model proposed by Vafai and Kim \cite{Vafai1989forced} lacks consistent validity across various parameters of fluid flow and porous media, including viscosity, permeability, and the inertial coefficient. Consequently, Nield et al. \cite{Nield1996forced} revisited the problem analyzed by Vafai and Kim \cite{Vafai1989forced} and presented a solution that remains uniformly valid for all the porous media parameter values.  The study by Nield et al. \cite{Nield1996forced} is noteworthy for presenting an exact solution; however, their velocity expression requires finding the roots of a cubic equation and performing numerical integration of an elliptic integral, both of which cannot be evaluated explicitly. These difficulties become even more pronounced in a two-layer system comprising both fluid and porous regions, where the coupling between the free fluid and porous region further complicates analytical tractability. 
Jha and Kaurangini \cite{Jha2011approximate} provided an approximate solution based on iteration for the time-dependent Brinkman-Forchheimer equation within a channel completely filled with an isotropic porous medium. In his study, Hooman \cite{Hooman2008perturbation} presented a perturbation solution for fully developed forced convection in a porous duct using the Brinkman-Forchheimer model. The asymptotic solutions correspond well with the numerical solution for both large and small Darcy numbers. The perturbation solution is effective for solving problems accurately at both low and high Darcy numbers, whereas it may not provide accurate results for moderate Darcy numbers. The studies mentioned above indicate that researchers have shown significant interest in understanding fluid flow within channels that are partially or fully filled with porous medium based on Brinkman-Forchheimer extended Darcy's model. There have been a few attempts to study this phenomenon when the porous medium has anisotropic properties and the inertial effect is taken into account. Kim et al. \cite{Kim2001effect} examined the influence of anisotropic permeability and diffusivity on the thermal characteristics of an aluminum foam heat sink, considering the inertial effect in the momentum equation. Nakayama et al. \cite{Nakayama2002heat} investigated fluid flow and heat transfer in an anisotropic porous medium to obtain the Forchheimer drag based on its anisotropic properties. Considering the inertial effect, Karmakar et al. \cite{Karmakar2019forced} studied forced convection in a channel filled with an anisotropic porous medium. They demonstrate that the heat transfer rate can significantly differ from an isotropic situation. Karmakar et al. \cite{Karmakar2022analysis} presented the existence and uniqueness results of Couette flow within a porous channel. In addition to these theoretical findings, they provided an approximate solution that is valid across a wider range of porous media parameter values. In a recent article, Karmakar et al. \cite{Karmakar2023couette} examined Couette-Poiseuille flow in a fluid overlying an anisotropic porous layer. They established results on both the existence and uniqueness of solutions, complemented by the numerical solutions.

Determining the velocity and shear stress distribution under negligible Forchheimer drag is a relatively straightforward task. However, when considering the impact of form drag by incorporating Forchheimer drag, i.e., a non-linear term in the momentum equation, the problem becomes more complex \cite{Hamdan1991analysis}. In this case, a uniformly valid exact solution is no longer feasible, and numerical integration becomes necessary. Asymptotic solutions are available for specific ranges of Darcy numbers (e.g., see Hooman \cite{Hooman2008perturbation}) but their applicability beyond these ranges necessitates additional validation. In recent years, artificial neural networks (ANNs) have gained popularity as a powerful statistical method for identifying relationships between variables. A neural network's ability to perform computation is based on the hope that we can reproduce some of the power and capability of the human brain by artificial means. Inspired by the biological neuron systems, artificial neural networks are constructed using a computer-based architecture that features several parallel components, including neurons and layers. These systems are composed of countless interconnected neurons that work together to address various problems. Learning in biological systems occurs by adjusting the synaptic connections among neurons. In contrast to that, ANNs primarily learn from the examples provided in the data, aiming to emulate this structure. The surge of interest in artificial neural networks was dramatically fueled after the invention of the backpropagation (BP) algorithm. Backpropagation is a supervised learning algorithm used to efficiently train multilayer artificial neural networks. Backpropagation is fundamental to the learning process of most artificial neural networks (ANNs), enabling them to adjust their internal parameters to improve prediction accuracy. Due to its strong mathematical foundation and potential applications, it is widely implemented and has proven effective in practice. In 1974, Paul Werbos first described this method \cite{Werbos1974beyond}. Later, Rumelhart, Hinton, and Williams \cite{Rumelhart1985learning} provided a clear and concise explanation of the BP algorithm. The artificial Levenberg-Marquardt method using a backpropagated neural network (ALMM-BNN) is a popular novel approach for handling various fluid flow problems  numerically, since it provides faster convergence and stability \cite{Afridi2025artificial,Shoaib2021intelligent}.

In this article, we examine a two-layer model enclosed by two parallel impermeable plates. The lower layer is an anisotropic Brinkman-Forchheimer-Darcy porous medium, while the upper layer is made up of a clear fluid with depth-dependent viscosity. The permeabilities of the porous medium vary along the flow direction, often described in the literature using the term anisotropic angle. In the present study, the upper plate moves with a constant velocity, while the lower plate remains stationary. At the liquid-porous interface, we take into account the Ochoa-Tapia and Whitaker \cite{Ochoa1995momentum} boundary condition, which necessitates a discontinuity in tangential stress while maintaining velocity continuity.  The current model provides valuable insights into the flow dynamics within and above the endothelial glycocalyx. It predicts the distribution of shear stress on the glycocalyx surface while considering variations in blood viscosity and the anisotropic properties of endothelial glycocalyx, while incorporating the effect of fluid inertia. This study seeks to predict the solution of the hydrodynamic problem using the Artificial Levenberg-Marquardt method combined with a backpropagated neural network (ALMM-BNN) paradigm for a moderate range of Darcy numbers when the typical perturbation methods fail to provide a correct result. We obtain the training dataset based on the asymptotic solution described in Section \ref{asymptotic solution}. Additionally, we propose an approximate solution that demonstrates good agreement with both the predicted and numerical solutions across a wider range of porous media parameter values. As per the author's knowledge, no prior investigation has paid attention so far to the impact of inertial effects on the hydrodynamic velocity and shear stress distribution in the presence of depth-dependent viscosity within a fluid layer overlying an anisotropic porous medium using the ALMM-BNN paradigm. Therefore, we initiate our investigation with a fundamental hydrodynamic analysis.

The layout of the manuscript is as follows: In Sec. \ref{Section 2}, we provide a broad description and show the mathematical formulation of the problem. In Sec. \ref{Section 3}, we introduce the non-dimensional variable and boundary conditions. A detailed analysis of asymptotic solutions obtained based on the perturbation method and approximate solutions obtained based on the iterative scheme is presented in Sec. \ref{asymptotic solution} and Sec. \ref{solution iterative based}, respectively. Section \ref{reults and discussion} comprises numerical and asymptotic results with the validation of the ALMM-BNN method. We explore the impact of several parameters, including the anisotropic ratio, orientation angle, Forchheimer number, and viscosity parameters, on the distribution of velocity and shear stress. As an application of our findings, we present key insights regarding the shear stress distribution on the arterial wall during blood flow within the artery.


\section{Mathematical formulation}\label{Section 2}

\begin{figure}[htp!]
\begin{center}
{\includegraphics[height=2.5 in, width=5.0 in]{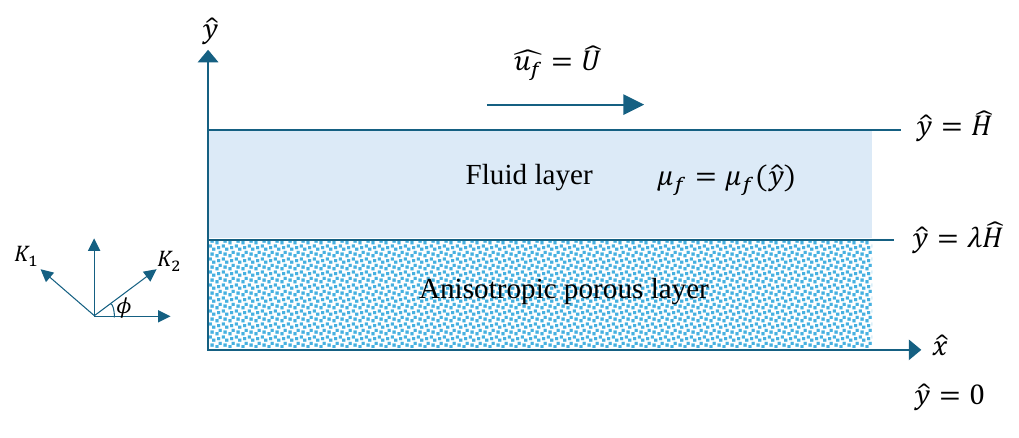}}\hspace{0.1in}
\caption{Schematic diagram of the physical situation}\label{Geometry}
\end{center}
\end{figure}

The topic under study is depicted schematically in Fig. \ref{Geometry}. It consists of a parallel plate channel partially filled with a porous layer having a depth-dependent viscosity overlying a porous layer. We assume that the porous medium is anisotropic in nature, characterized by constant permeabilities $K_{1}$ and $K_{2}$ along the two principal axes. The orientation angle, known as the anisotropic angle, is defined as the angle between the horizontal direction and the principal axis associated with permeability $K_{2}$. We begin by outlining the generic form of the governing equations in vector form for each layer. The flow inside the fluid layer is governed by the equation of continuity along with the Navier-Stokes equations as follows.

$\bullet$ \textbf{Fluid layer:} ($\lambda H \leq \hat{y} \leq H$)
\begin{equation}\label{eq1}
\hat{\nabla}.\hat{\mathbf{V}_{f}}=0,    
\end{equation}
\begin{equation}\label{eq2}
\left(\hat{\mathbf{V}_{f}}\cdot \hat{\nabla}\right)\hat{\mathbf{V}_{f}}=-\frac{1}{\rho_{f}} \hat{\nabla}\hat{P}_{f} +\frac{1}{\rho_{f}}  \hat{\nabla}.\left(\mu_{f}\left(\hat{\nabla}\hat{\mathbf{V}_{f}}+(\hat{\nabla}\hat{\mathbf{V}_{f}})^{T}\right)\right),   
\end{equation}
where $\hat{\mathbf{V}_{f}}=(\hat{u_{f}},\hat{v_{f}})$ is the dimensional velocity vector, $\rho_{f}$ is the density, $\hat{P_{f}}$ is the pressure, and $\mu_{f}$ is the viscosity. 

The corresponding mass and momentum equation in the porous layer is given below.

$\bullet$ \textbf{Porous layer:} ($0\leq \hat{y}\leq \lambda H$)
\begin{equation}\label{eq3}
 \hat{\nabla}.\hat{\mathbf{V}_{p}}=0,   
\end{equation}
\begin{equation}\label{eq4}
-\hat{\nabla} \hat{P}_{p}+ \mu_{\textrm{eff}} \hat{\nabla}^{2} \hat{\mathbf{V}_{p}}-\mu \mathbf{\overline{K}}^{-1}\hat{\mathbf{V}_{p}}- c_{F}\rho_{f}\mathbf{\overline{K}}^{-1/2} |\hat{\mathbf{V}_{p}}| \hat{\mathbf{V}_{p}}=0,   
\end{equation}
where $\hat{\mathbf{V}_{p}}=(u_{p},v_{p})$ is the velocity vector, $P_{p}$ is the pressure, $\mu_{\textrm{eff}}$ is the effective viscosity, and $c_{F}$ is the inertial coefficient. Since the porous medium is anisotropic in nature, the permeability is represented by a second-rank tensor given by
\begin{equation}\label{eq5}
\mathbf{\overline{K}}=
\begin{bmatrix}
K_{1}\sin^{2}\phi+ K_{2} \cos^{2} \phi & \left(K_{2}-K_{1}\right) \sin \phi \cos \phi \\
\left(K_{2}-K_{1}\right) \sin \phi \cos \phi & K_{2}\sin^{2}\phi+ K_{1} \cos^{2} \phi
\end{bmatrix}.   
\end{equation}
Eq. (\ref{eq4}) is commonly known as the Brinkman-Forchheimer-extended Darcy's equation.

\begin{thm}\label{horn}
(see p. 439 in \cite{Horn2012matrix}) Let $A$ be a $n\times n$ Hermitian and positive definite matrix, and let 
$m\in \{2,3,4...\},$ then there exists a unique Hermitian positive semidefinite matrix $B$ such that $B^{m}=A$.
\end{thm}
Clearly, the matrix $\mathbf{\overline{K}}$ in Eq.~(\ref{eq5}) is symmetric with eigenvalues $K_{1}$ and $K_{2}$. Since, $K_{1}, K_{2} > 0$, $\mathbf{\overline{K}}$ is positive definite. By Theorem \ref{horn}, one may get a unique positive semidefinite matrix $B$ such that $B^2=\mathbf{\overline{K}}$. Hence, $\mathbf{\overline{K}}^{-1/2}$ exists and is positive semidefinite and unique. 

\subsection{Assumptions on the current model and boundary conditions} 
\begin{itemize}
    \item The fluid and porous zone exhibits fully developed, unidirectional flow (i.e., $\hat{v}_{f}=0, \hat{v}_{p}=0$). This indicates that the axial ($x$-direction) velocity is solely determined by the transverse ($y$-direction) coordinate, as the continuity equation requires. 
    \item We assume that the fluids in both layers are Newtonian. The viscosity of the fluid layer has a depth dependency and equals the fluid's viscosity in the porous layer at the interface. In this work, we assume
    \begin{equation}\label{eq5_viscosity}
      \mu_{f}=\mu\left[1+A (\hat{y}-\lambda H)\right].  
    \end{equation}
    \item We assume that the upper plate $\hat{y}=H$ is moving with a constant velocity $\hat{U}$ in the horizontal direction in its plane, and the lower plate at $\hat{y}=0$ is stationary. Thus, we get
    \begin{equation}\label{eqbcs1}
        \hat{u}_{f}=\hat{U} \quad \textrm{at} \quad \hat{y}=H, ~\textrm{and} \quad \hat{u}_{p}=0   \quad \textrm{at} \quad \hat{y}=0.
    \end{equation}
    \item We use the continuity of velocity at the interface and Ochoa-Tapia stress jump condition \cite{Ochoa1995momentum,Ruan2026stokes,Alazmi2001analysis}, i.e., 
    \begin{equation}\label{eqbcs3}
        \hat{u}_{f}=\hat{u}_{p} \quad \textrm{at} \quad \hat{y}=\lambda H,
    \end{equation}
    \begin{equation}\label{eqbcs4}
      \mu_{\textrm{eff}} \frac{d \hat{u}_{p}}{d \hat{y}}- \mu_{f}(y) \frac{d \hat{u}_{f}}{d \hat{y}}=\mu \beta \frac{\hat{u}_{p} b}{\sqrt{K_{1}}}.
       \end{equation}
       Here $\beta$ is the stress jump parameter. If $\beta=0$, one gets the continuity of the tangential stress at the interface. 
\end{itemize}
Under these assumptions, the governing equations for the fluid and porous regions are given by:

$\bullet$ \textbf{Fluid layer} ($\lambda H \leq \hat{y} \leq H$):
\begin{equation}\label{eq6}
\frac{\partial \hat{u}_{f}}{\partial \hat{x}}=0,    
\end{equation}
\begin{equation}\label{eq7}
-\frac{1}{\rho_{f}} \frac{\partial \hat{P}_{f}}{\partial \hat{x}}+ \frac{1}{\rho_{f}} \frac{d}{d\hat{y}}\left(\mu_{f}(\hat{y}) \frac{d \hat{u}_{f}}{d\hat{y}}\right)=0,
\end{equation}
\begin{equation}\label{eq8}
-\frac{1}{\rho_{f}}\frac{\partial \hat{P}_{f}}{\partial \hat{y}}=0.    
\end{equation}
$\bullet$ \textbf{Porous layer} ($0\leq \hat{y}\leq \lambda H$):
\begin{equation}\label{eq9}
\frac{\partial \hat{u}_{p}}{\partial \hat{x}}=0, 
\end{equation}
\begin{equation}\label{eq10}
 -\frac{\partial \hat{P}_{p}}{\partial \hat{x}}+ \mu_{\textrm{eff}} \frac{d^{2}\hat{u}_{p}}{d \hat{y}^{2}} -\frac{\mu a}{K_{1}}  \hat{u}_{p}-   \frac{\rho_{f} c_{F} b}{\sqrt{K_{1}}}\hat{u}_{p}^{2}=0, 
\end{equation}
\begin{equation}\label{eq11}
 -\frac{\partial \hat{P}_{p}}{\partial \hat{y}}-\frac{\mu c}{K_{1}} \hat{u}_{p}-\frac{\rho_{f}c_{F} d}{\sqrt{K_{1}}}  \hat{u}_{p}^{2}=0,   
\end{equation}
where $a=\sin^{2}\phi+K \cos^{2}\phi$, $b=\sin^{2}\phi+\sqrt{K} \cos^{2}\phi$, $c=(K-1)\sin \phi \cos \phi$, 
$d=(\sqrt{K}-1)\sin \phi \cos \phi$, are parameters characterizing the anisotropy and $K=K_{1}/K_{2}$ is the anisotropic permeability ratio.

From Eq. (\ref{eq8}), the pressure gradient is zero in the $\hat{y}$ direction. Consequently, $\frac{\partial \hat{P}_{f}}{\partial \hat{x}}=\frac{d\hat{P}_{f}}{d\hat{x}}=\textrm{constant}=-\hat{Q}~(\textrm{say}).$

Using Eq. (\ref{eq9}) in Eqs. (\ref{eq10}) and (\ref{eq11}), we get that 
\begin{equation}\label{eq12}
0=\frac{\partial}{\partial \hat{x}}\left(\frac{\partial \hat{P}_{p}}{\partial \hat{x}}\right) \quad \textrm{and} \quad 0= \frac{\partial}{\partial \hat{x}}\left(\frac{\partial \hat{P}_{p}}{\partial \hat{y}}\right),    
\end{equation}
i.e., 
\begin{equation}\label{eq13}
 0=\frac{\partial}{\partial \hat{x}}\left(\frac{\partial \hat{P}_{p}}{\partial \hat{x}}\right) \quad   \quad \textrm{and} \quad  0= \frac{\partial}{\partial \hat{y}}\left(\frac{\partial \hat{P}_{p}}{\partial \hat{x}}\right).
\end{equation}
Hence, it follows that $\frac{\partial \hat{P}_{p}}{\partial \hat{x}}=\textrm{constant}=-G~ (\textrm{say})$.

\section{Non-dimensionalization of governing equations and boundary conditions} \label{Section 3}
We introduce the following non-dimensional variables:
\begin{equation} \label{nondim1}
y=\frac{\hat{y}}{H}, \quad (u_{f},u_{p})=\frac{(\hat{u}_{f}, \hat{u}_{p})}{GH^{2}/\mu}, \quad M=\frac{\mu_{\textrm{eff}}}{\mu}, \quad Da=\frac{K_{1}}{H^{2}}, \quad F=\frac{c_{F}\rho_{f} G H^{3}}{\mu^{2}},    
\end{equation}
where $M$ represents the viscosity ratio, $Da$ denotes the Darcy number, and $F$ is the Forchheimer number indicating the inertial drag. 

On non-dimensionalization the governing equations in the fluid and porous regions are

$\bullet$ \textbf{Fluid layer} ($\lambda\leq y\leq 1$):
\begin{equation}\label{nondim2}
 \frac{d}{dy}\left(\left(1+f(y-\lambda)\right) \frac{du_{f}}{dy})\right)+Q=0,   
\end{equation}
where $Q=\hat{Q}/G$ is the non-dimensional pressure gradient and $f=AH$. 

$\bullet$ \textbf{Anisotropic porous layer} ($0\leq y \leq \lambda$):
\begin{equation}\label{nondim3}
 1+M \frac{d^{2}u_{p}}{dy^{2}}-\epsilon^{2} au_{p}-F b \epsilon u_{p}^{2}=0,   
\end{equation}
where $\epsilon=1/\sqrt{Da}$ is the porous media shape parameter, a non-dimensional quantity. 

The boundary conditions given in Eqs. (\ref{eqbcs1})-(\ref{eqbcs4}) in nondimensional form are given as
\begin{equation}\label{nondim4}
 u_{f}=U, \quad \textrm{at} \quad y=1.   
\end{equation}
At $y=\lambda$, 
\begin{equation}\label{nondim5}
 u_{f}=u_{p}, \quad M \frac{du_{p}}{dy}-\left[1+ f (y-\lambda)\right] \frac{du_{f}}{dy}=\frac{\beta d}{\sqrt{Da}}u_{p}.  
 \end{equation}
 At $y=0$, 
 \begin{equation}\label{nondim6}
  u_{p}=0.   
 \end{equation}

\section{Asymptotic solutions} \label{asymptotic solution}
For the asymptotic analysis, we assume that $F, ~M,~ K \sim O(1).$

\subsection{Solution for large Darcy number} \label{large Darcy}
When the Darcy number is large, we have $Da\gg 1$, i.e., $\epsilon\ll 1$. We seek an asymptotic expansion of the velocity distribution as follows:
\begin{equation}\label{eq_expansion}
 \left(u_{f},u_{p}\right)=\left(u_{f}^{(0)},u_{p}^{(0)}\right)+\epsilon \left(u_{f}^{(1)},u_{p}^{(1)}\right)+ \epsilon^{2} \left(u_{f}^{(2)},u_{p}^{(2)}\right) + O(\epsilon^{3}).  
\end{equation}

Equation (\ref{nondim2}) and expansion (\ref{eq_expansion}) provide the leading, first-order, and second-order terms as follows:
\begin{equation}\label{largeda2}
O(1): \frac{d}{dy} \left(\left(1+ f (y-\lambda)\right) \frac{du_{f}^{(0)}}{dy}\right)+Q=0,    
\end{equation}
\begin{equation}\label{largeda3}
O(\epsilon): \frac{d}{dy} \left(\left(1+ f (y-\lambda)\right) \frac{du_{f}^{(1)}}{dy}\right)=0,    
\end{equation}
\begin{equation}\label{largeda4}
O(\epsilon^{2}): \frac{d}{dy} \left(\left(1+ f (y-\lambda)\right) \frac{du_{f}^{(2)}}{dy}\right)=0.    
\end{equation}

In a similar manner, Equation (\ref{nondim3}) and the expression (\ref{eq_expansion}) yield the following leading, first-order, and second-order terms, respectively.

\begin{equation}\label{largeda5}
O(1): M \frac{d^{2}u_{p}^{(0)}}{dy^{2}}+1=0,
\end{equation}
\begin{equation}\label{largeda6}
 O(\epsilon): M \frac{d^{2}u_{p}^{(1)}}{dy^{2}}- F b (u_{p}^{(0)})^{2}=0,   
\end{equation}
\begin{equation}\label{largeda7}
O(\epsilon^{2}): M \frac{d^{2}u_{p}^{(2)}}{dy^{2}}- a u_{p}^{(0)}- 2 F b u_{p}^{(0)} u_{p}^{(1)}=0.
\end{equation}

The boundary conditions are also expanded with the help of Eq.~(\ref{eq_expansion}).
\begin{equation}\label{largeda8}
u_{f}^{(0)}=-\frac{Q y}{f}+\ln\left(1+f(y-\lambda)\right) \left(-\frac{Q\lambda}{f}+\frac{D_{10}}{f}+\frac{Q}{f^{2}}\right)+D_{20},
\end{equation}
\begin{equation}\label{largeda9}
u_{p}^{(0)}=-\frac{y^{2}}{2 M}+ C_{10}y+C_{20},    
\end{equation}
\begin{equation}\label{largeda10}
 u_{f}^{(1)}=D_{11} \ln\left(y-\frac{(f \lambda-1)}{f}\right)+D_{21},   
\end{equation}
\begin{multline}\label{largeda11}
 u_{p}^{(1)}=\frac{1}{4M^{3}} \left(Fb \left(\frac{y^{6}}{30}-\frac{C_{10}M}{5}y^{5}+\frac{(20C_{10}^{2} M^{2}-20 C_{20}M)}{60} y^{4}+\frac{4 C_{10} C_{20} M^{2}}{3}y^{3}\right.\right. \\\left.\left.+2 C_{20}^{2} M^{2} y^{2} \right)\right)+C_{11}y+C_{21},   
 \end{multline}
\begin{equation}\label{largeda12}
 u_{f}^{(2)}=D_{12}\ln\left(y-\frac{(f \lambda-1)}{f}\right)+D_{22},  
\end{equation}
\begin{multline}\label{largeda13}
u_{p}^{(2)}(y)=-\frac{F^{2} b^{2} y^{10}}{10800 M^{5}}+\frac{\mathit{C10} \ F^{2} b^{2} y^{9}}{1080 M^{4}}+\frac{\mathit{C20} \ F^{2} b^{2} y^{8}}{560 M^{4}}- \frac{11 \mathit{C10}^{2} F^{2} b^{2} y^{8}}{3360 M^{3}}\\-\frac{\mathit{C10} \mathit{C20}  F^{2} b^{2} y^{7}}{70 M^{3}} +\frac{\mathit{C10}^{3} F^{2} b^{2} y^{7}}{252 M^{2}}+\frac{\mathit{C20} \ \mathit{C10}^{2} F^{2} b^{2} y^{6}}{36 M^{2}}-\frac{\mathit{C20}^{2} F^{2} b^{2} y^{6}}{45 M^{3}}\\+\frac{\mathit{C10} \ \mathit{C20}^{2} F^{2} b^{2} y^{5}}{12 M^{2}}-\frac{\mathit{C11} F \ y^{5} b}{20 M^{2}}+\frac{\mathit{C10} \mathit{C11} F \ y^{4} b}{6 M}-\frac{\mathit{C21} F \ y^{4} b}{12 M^{2}}\\+\frac{\mathit{C20}^{3} F^{2} b^{2} y^{4}}{12 M^{2}}-\frac{y^{4} a}{24 M^{2}}+\frac{\mathit{C10} \mathit{C21} F \ y^{3} b}{3 M}+\frac{\mathit{C11} \mathit{C20} F \ y^{3} b}{3 M}+ \frac{\mathit{C10} \ y^{3} a}{6 M}\\+\frac{\mathit{C20} \mathit{C21} F \ y^{2} b}{M}+\frac{\mathit{C20} \ y^{2} a}{2 M}+\mathit{C12} y +\mathit{C22},    
\end{multline}
where $C_{10},C_{20}, D_{10}, D_{20}, C_{11}, C_{21}, D_{11}, D_{21}, C_{12}, C_{22}, D_{12}, D_{22}$ are arbitrary constants those can be obtained by a software packages such as MATHEMATICA.

\subsection{Solution for small Darcy number}\label{solution small Darcy}
When the Darcy number is small enough, we have $Da\ll 1.$ Accordingly, $\epsilon^{2}=1/Da$ leads to $1/\epsilon \ll 1.$ 
Eq. (\ref{nondim3}) correspondingly takes the following form:
\begin{equation}\label{smallda1}
\frac{M}{\epsilon^{2}} \frac{d^{2}u_{p}}{dy^{2}}-a u_{p}-\frac{F b}{\epsilon}u_{p}^{2}+\frac{1}{\epsilon^{2}}=0.    
\end{equation}
We use the stretching principle to address this singular perturbation problem, which is caused by the highest-order derivative being multiplied by a small parameter $1/\epsilon^{2}$. 
The outer solution is given by 
\begin{equation}\label{smallda2}
 u_{p}^{\textrm{out}}=\frac{1}{2}\left(-\frac{a\epsilon}{Fb}+\sqrt{\left(\frac{a \epsilon}{Fb}\right)^{2}+\frac{4}{F b \epsilon}} \right).  
\end{equation}

By introducing the stretching variable $\xi=\epsilon y$, we can get the inner solution. The corresponding equation then reduces to 
\begin{equation}\label{smallda3}
\frac{d^{2}u_{p}^{\textrm{in}}}{d\xi^{2}}-\frac{a}{M} u_{p}^{\textrm{in}}=0.    
\end{equation}
The composite solution is obtained using the Prandtl matching principle \cite{Bush2018perturbation} as 
\begin{equation}\label{smallda4}
u^{\textrm{comp}}=u^{\textrm{out}}+u^{\textrm{in}}-\left(u^{\textrm{in}}\right)^{\textrm{out}}.    
\end{equation}
Therefore, 
\begin{equation}\label{smallda5}
u^{\textrm{comp}}= u_{p}^{\textrm{out}}\left(1-e^{-\sqrt{\frac{a}{M}}\xi}\right),  
\end{equation}
yields the composite solution that is uniformly valid throughout the region. 
The corresponding solution in the fluid region is given by 
\begin{equation}\label{smallda6}
u_{f}(y)=-\frac{Qy}{f}-\frac{\ln(f(y-\lambda)+1)}{f}\left(Q\lambda-T_{1}-\frac{Q}{f}\right)+T_{2},   
\end{equation}
where the constants $T_{1}$ and $T_{2}$ can be obtained using the continuity of velocities at the interface and the condition $u_{f}(1)=U$.
\begin{equation}\label{smallda7}
T_{1}=\frac{e^{-\sqrt{\frac{a}{M}}} \epsilon \lambda f^{2} u^{\textrm{out}}+ Q (f \lambda -1) \ln(-f \lambda+f+1)-f((-U+u^{\textrm{out}})f+ Q(\lambda-1))}{\ln(-f \lambda+f+1) f},
\end{equation}
\begin{equation}\label{smallda8}
T_{2}=-\frac{(u^{\textrm{out}} e^{-\sqrt{\frac{a}{M}}} \epsilon \lambda f- Q \lambda-u^{\textrm{out}}f)}{f}.    
\end{equation}

\section{Proposed approximate solution} \label{solution iterative based}
In this section, we propose an approximate solution in addition to the asymptotic solution. To do this, we linearize the nonlinear Forchheimer term by assuming $P=Fb \epsilon u_{p}$. This choice reduces Eq.~(\ref{nondim3}) to an auxiliary linear model as 
\begin{equation}\label{approximate1}
M \frac{d^{2}u_{p}}{dy^{2}}-\epsilon^{2} a u_{p}-P u_{p}+1=0,
\end{equation}
that admits a solution in the classical sense given by
\begin{equation}\label{approximate2}
u_{f}(y)=-\frac{Qy}{f}-\frac{\ln(f(y-\lambda)+1)}{f}\left(Q\lambda-E_{1}-\frac{Q}{f}\right)+E_{2},    
\end{equation}
\begin{equation}\label{approximate3}
u_{p}(y)=F_{2} e^{\sqrt{\frac{a\epsilon^{2}+P}{M}} y}  +F_{1}  e^{-\sqrt{\frac{a\epsilon^{2}+P}{M}} y} +\frac{1}{a\epsilon^{2}+P}.
\end{equation}
Therefore, with an aim to solve the nonlinear equation, we now define 
$$u_{p}(y)=\Phi(u), \quad u_{f}(y)=\Omega(u).$$ Correspondingly, we define the iterative process as 
\begin{equation}\label{approximate6}
 u_{p_{i+1}}=\Phi(u_{p_{i}}), \quad  u_{f_{i+1}}=\Omega(u_{f_{i}}).
\end{equation}
The iterative solution is obtained compared with that of previous iteration (initial guess $i=0$) and this process is continued until the maximum relative error in the values of local velocity between two successive iterations become less than $10^{-5}$, i.e., $|u_{f_{i+1}}-u_{f_{i}}|<10^{-5},\,\,\mbox{and}\,\, \quad |u_{p_{i+1}}-u_{p_{i}}|<10^{-5},$ where
\begin{equation}\label{approximate4}
u_{f_{0}}(y)= -\frac{Qy}{f}-\frac{\ln(f(y-\lambda)+1)}{f}\left(Q \lambda -D_{1}-\frac{Q}{f}\right)+D_{2},\,\,\mbox{and}   
\end{equation}
\begin{equation}\label{approximate5}
u_{p_{0}}(y)= C_{1}  e^{y\sqrt{\frac{a}{M Da}}}   + C_{2} e^{-y\sqrt{\frac{a}{M Da}}}+\frac{Da}{a}.
\end{equation}

\begin{figure}[h!]
    \centering
   {\subfigure[\label{phi_less_than1}]{\includegraphics[height=2 in, width=2.6 in]{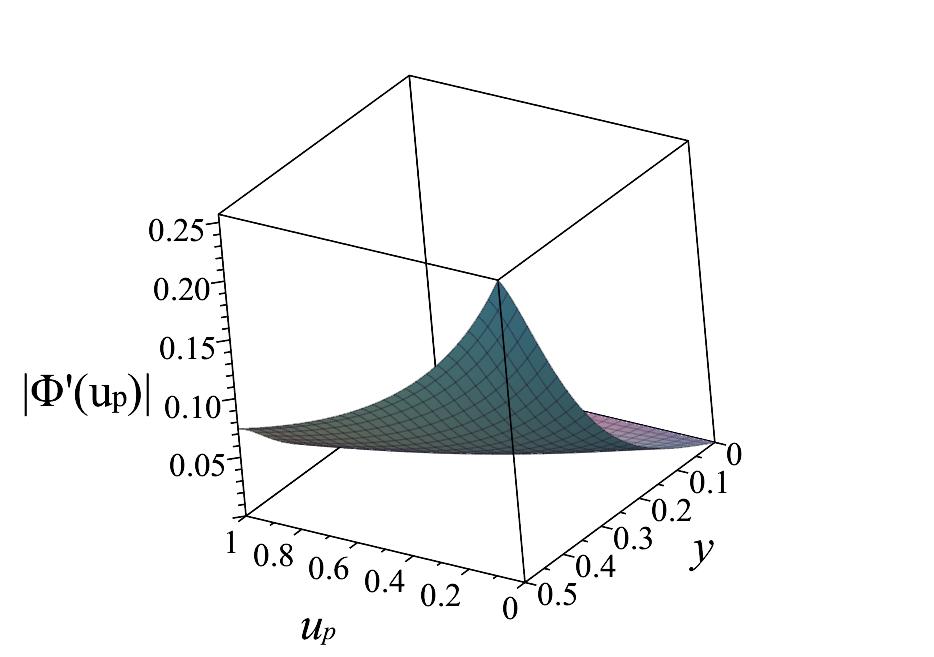}}}
    {\subfigure[\label{omega_less_than1}]{\includegraphics[height=2 in, width=2.6 in]{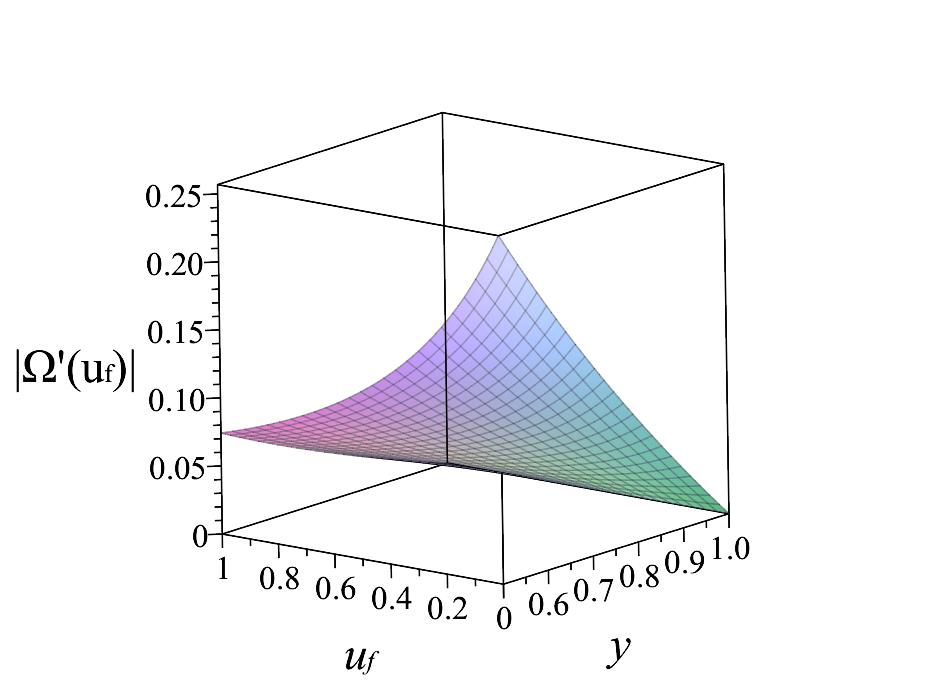}}}
    \caption{(a) $|\Phi'(u_{p})|$ behavior for $0\leq y \leq \lambda$; (b) $|\Omega'(u_{f})|$ behavior for $\lambda\leq y \leq 1$; when $Q=1, M=1, f=1, \beta=0.5, K=1, \phi=0, \lambda=0.5, F=10, Da=0.01,$ and $U=1.$}
    \end{figure}

To demonstrate the convergence of the iterative scheme given in Eq. (\ref{approximate6}), we have plotted the variation of $\Phi'(u_{p})$ and $\Omega'(u_{f})$ inside the channel while considering $0\leq u_{f} \leq 1$, and $0 \leq u_{p} \leq 1$. The reason behind such a range of $u_{f}$ and $u_{p}$ is due to the maximum upper plate velocity being set to unity in this problem. In Figs. \ref{phi_less_than1} and \ref{omega_less_than1}, we observe that $|\Phi'(u_{p})|<1$ and $|\Omega'(u_{f})|<1$, those satisfy the convergence condition of Eq.~(\ref{approximate6}). We have also verified the convergence conditions for other relevant parameter combinations listed in Table \ref{tab:parameter}, although those details are not included due to space constraints.  

\section{Results and discussion} \label{reults and discussion}
In Sec.~\ref{asymptotic solution} and Sec.~\ref{solution iterative based}, we present the asymptotic solutions and the iterative approximate solutions, respectively. The momentum equations given in Eq.~(\ref{nondim2}) and Eq.~(\ref{nondim3}) form a system of second-order ordinary differential equations. Eq.~(\ref{nondim3}) is a nonlinear differential equation that makes the system analytically untractable. The simplified model equations and their associated interface and boundary conditions do not allow for an analytical solution. Hence, one has to rely on a numerical solution. Numerical results are generated with the help of MATLAB 2023a, utilizing the $\textrm{bvp5c}$ function for solving the multipoint boundary value problem. For numerical simulations, we used specific parameter ranges as presented in Table.~\ref{tab:parameter}.   

\begin{table}[h!]
    \centering
    \begin{tabular}{p{1.8 cm}p{5.5 cm}p{3 cm}p{1.6 cm}}
       Parameters  & Description & Value & References  \\
       \hline
        $Da$ & Non-dimensional permeability & $10^{-4}\leq Da \leq 10$ & \cite{Nield1996forced,Karmakar2019forced}\\ 
        $F$ & Non-dimensional inertial coefficient & $0\leq F\leq 100$ &\cite{Hooman2008perturbation,Karmakar2019forced}\\
        $M$ & Viscosity ratio & $0.5\leq M \leq 10$ &\cite{Nield1996forced,Givler1994determination}\\
        $K$ & Anisotropic permeability ratio & $0.5\leq K\leq 4$ & \cite{Karmakar2022analysis,Karmakar2019forced}\\ 
        $\phi$ & Orientation angle & $0\leq \phi \leq \frac{\pi}{2}$ &\cite{Karmakar2022analysis,Karmakar2019forced}\\ 
        \hline
    \end{tabular}
    \caption{List of parameters range used for numerical simulation}
    \label{tab:parameter}
\end{table}

\begin{figure}[h!]
    \centering
   {\subfigure[\label{velocity low Da}]{\includegraphics[width=2.6 in]{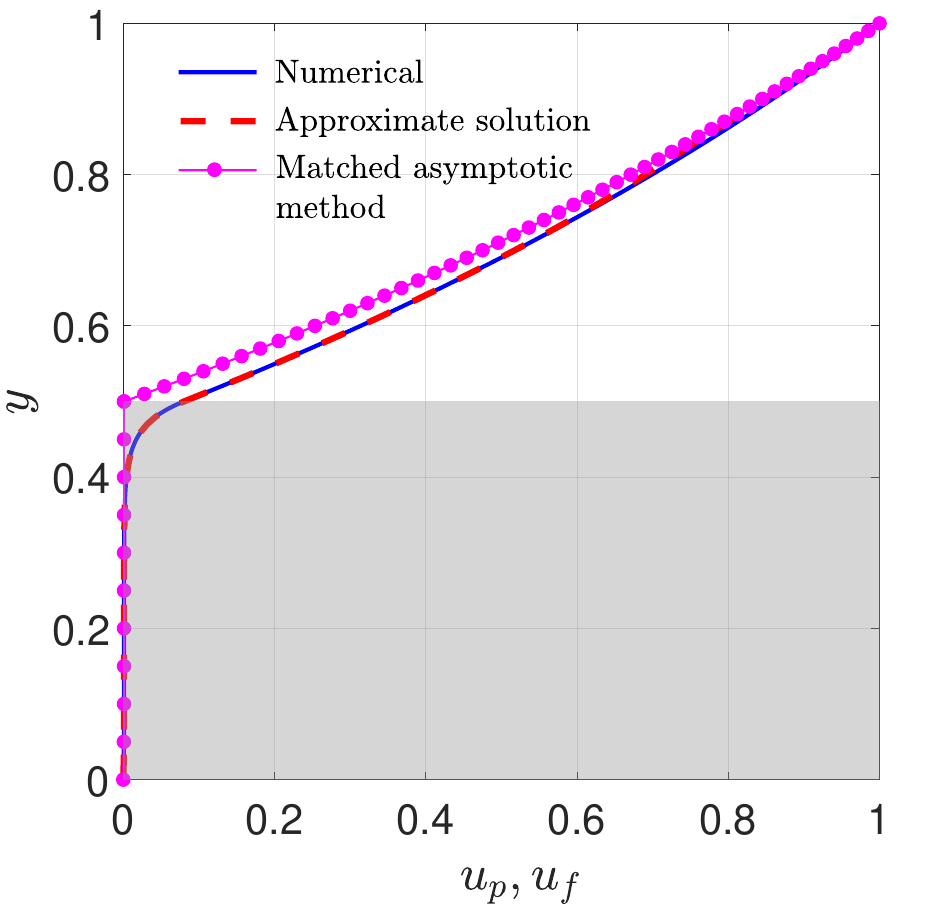}}}
    {\subfigure[\label{velocity high Da}]{\includegraphics[width=2.6 in]{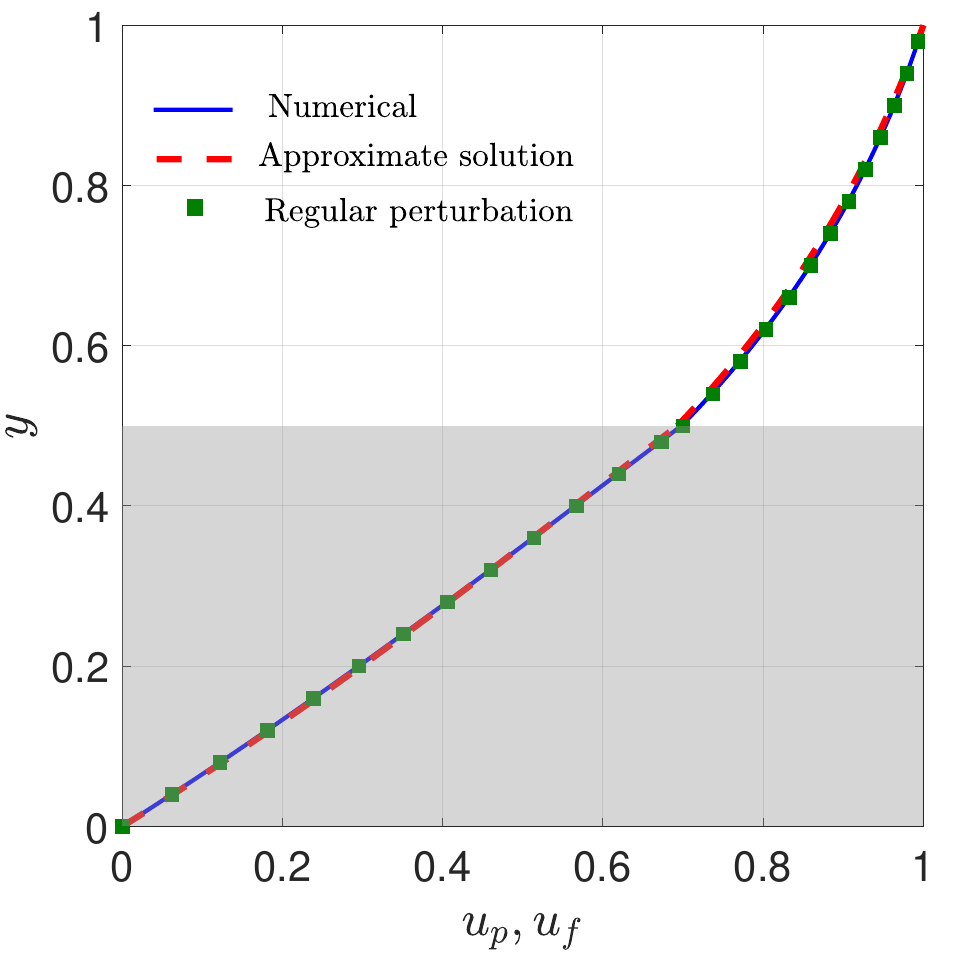}}}
    \caption{Dimensionless velocity profile (a) for $\lambda=0.5$, $Q=1$, $M=1$, $\beta=0$, $Da=0.001$, $K=1$, $F=1$, $U=1$, $f=1$; (b) for $\lambda=0.5$, $Q=1$, $M=1$, $\beta=0.5$, $Da=1$, $K=1$, $F=1$, $U=1$, $f=1$.}
    \label{Velocity low and high Da}
\end{figure}
In Fig. \ref{Velocity low and high Da}, we present the asymptotic results alongside those obtained from the numerical computations. Fig. \ref{velocity low Da} illustrates the velocity distribution for low values of $Da$. For low values of $Da$, the corresponding problem is a singular perturbation problem, which is addressed with the help of matched asymptotic expansion. For high values of $Da$, the velocity distribution is derived from a regular perturbation analysis, which is presented in Fig. \ref{velocity high Da}. In both the cases, we see that the asymptotic solutions and proposed approximate solutions align well with the numerical results. Although asymptotic methods are effective for very high or very low values of the Darcy number, they tend to lose accuracy in the moderate range, limiting their applicability in such cases. In this work, we have developed an approximate solution using an iterative scheme that remains uniformly valid over a broader range of parameters and demonstrates good agreement with the numerical results. 

\begin{figure}[h!]
    \centering
   {\subfigure[\label{velocity different K}]{\includegraphics[width=2.6 in]{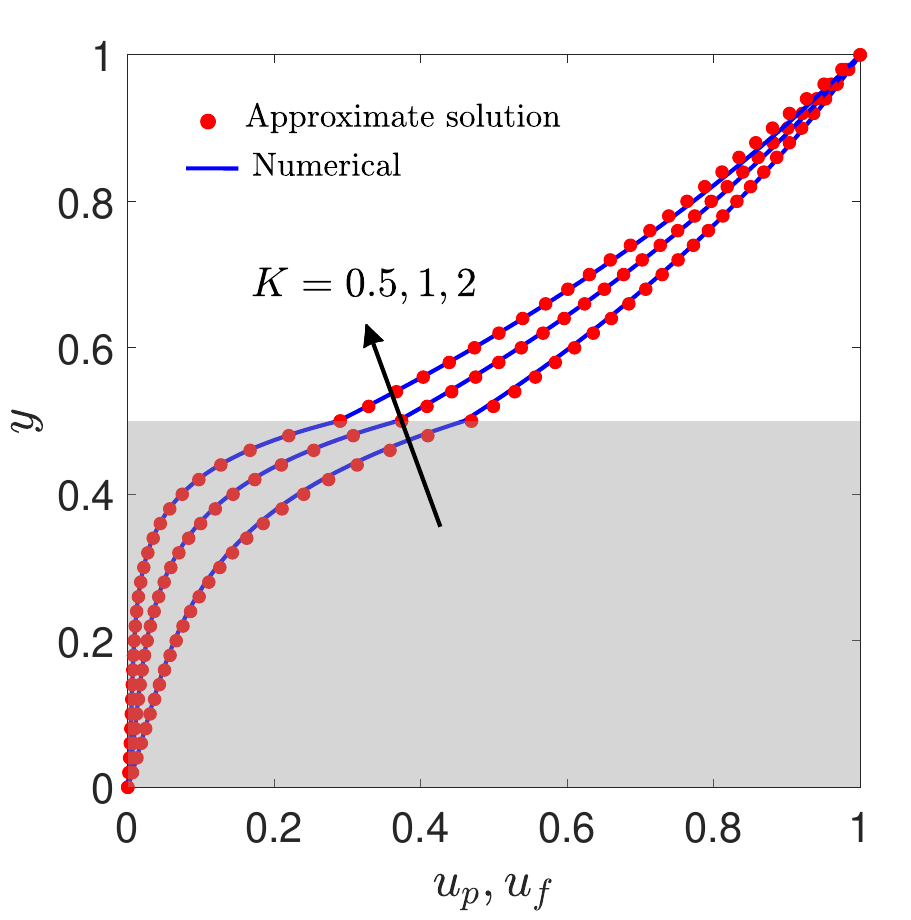}}}
    {\subfigure[\label{velocity different phi}]{\includegraphics[width=2.6 in]{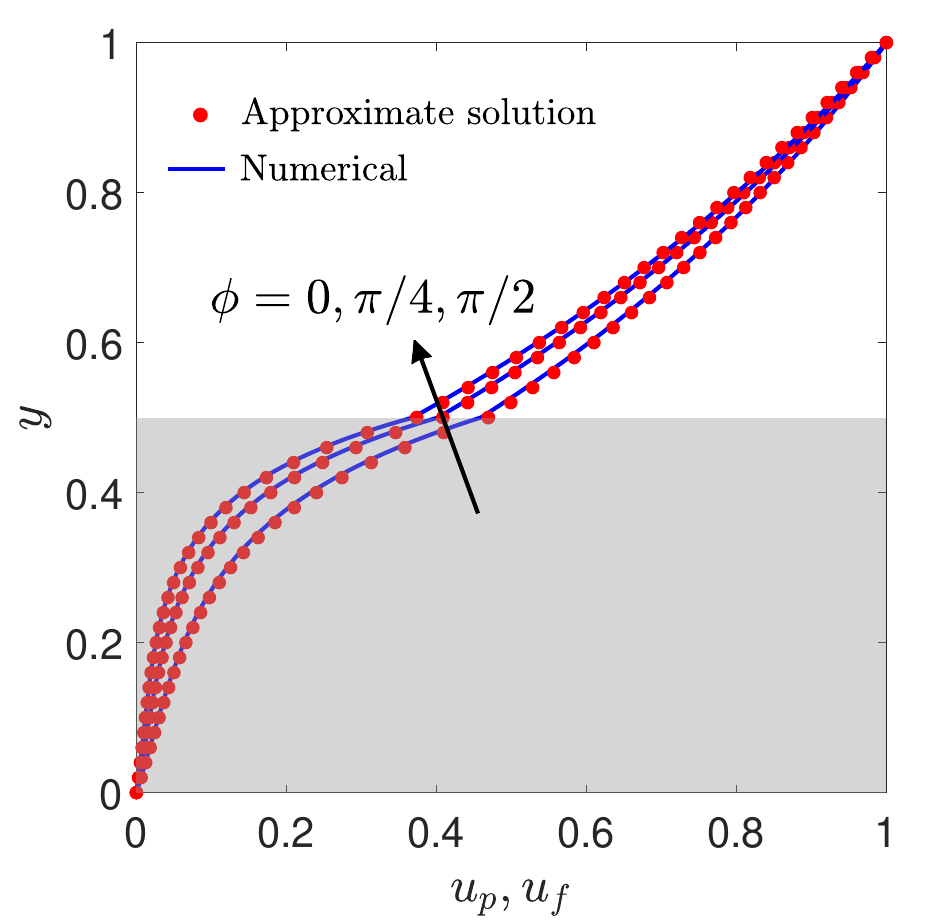}}}
    \caption{Dimensionless velocity profile (a) for different $K$ when $\lambda=0.5$, $Q=1$, $M=1$, $\beta=0.5$, $Da=0.01$, $F=1$, $U=1$, $f=1$, and $\phi=0$; (b) for different $\phi$ when $\lambda=0.5$, $Q=1$, $M=1$, $\beta=0.5$, $Da=0.01$, $K=0.5$, $F=1$, $U=1$, and $f=1$.}
   \end{figure}

Fig~\ref{velocity different K} depicts the horizontal velocity distribution across the channel for different anisotropic ratios $K$. The upper plate moves with a non-dimensional velocity, $U=1$, which is the maximum fluid velocity. As the flow enters the porous medium, the resistance introduced by the porous packings causes the velocity to decrease progressively and ultimately drop to zero at the stationary lower plate owing to the no-slip boundary condition. For a fixed value of $Da<1$, an increase in $K$ leads to a reduction in velocity. This behavior aligns with the fact that when $\phi=0$, and $Da<1$ (i.e., ${K_1}$ is fixed), increasing $K$ leads to a reduced permeability ${K_2}$ in the horizontal direction, which causes the velocity to decrease. The influence of $\phi$, the angle between the horizontal direction and the principal axis associated with permeability $K_{2}$, is depicted in Fig.~\ref{velocity different phi} for different values. As the orientation angle $\phi$ increases, the associated velocity decreases. This is because for a fixed $Da$ (i.e., $K_{1}$), when $K<1$, setting $\phi=0$ aligns the higher permeability axis with the horizontal (flow) direction, enhancing the fluid movement along the flow. In contrast, when $K>1$, the configuration $\phi=\pi/2$ aligns the more permeable axis with the flow direction, and it enhances momentum transport (for brevity, results are not shown here). In both Figs.~\ref{velocity different K} and \ref{velocity different phi}, the obtained approximate solutions show a good agreement with the numerical result. 

\begin{figure}[h!]
    \centering
   {\subfigure[\label{velocity different Da}]{\includegraphics[width=2.7 in]{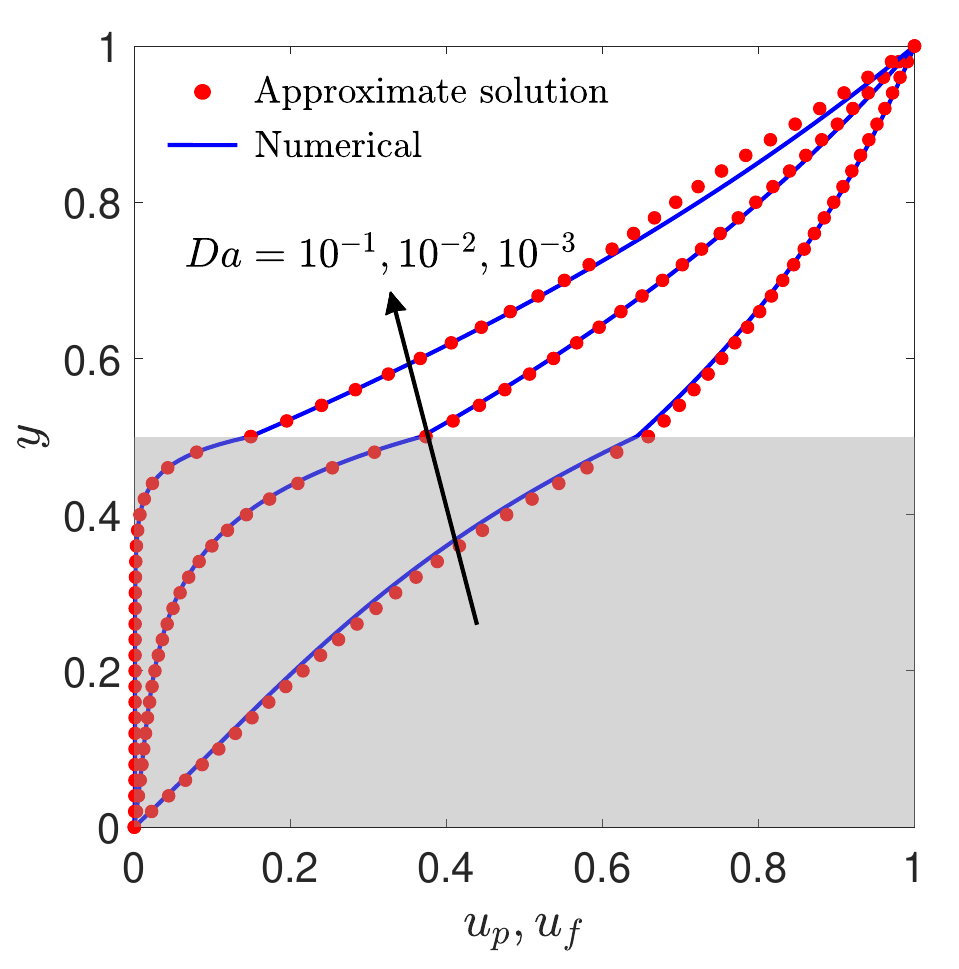}}}
    {\subfigure[\label{velocity different F}]{\includegraphics[width=2.6 in]{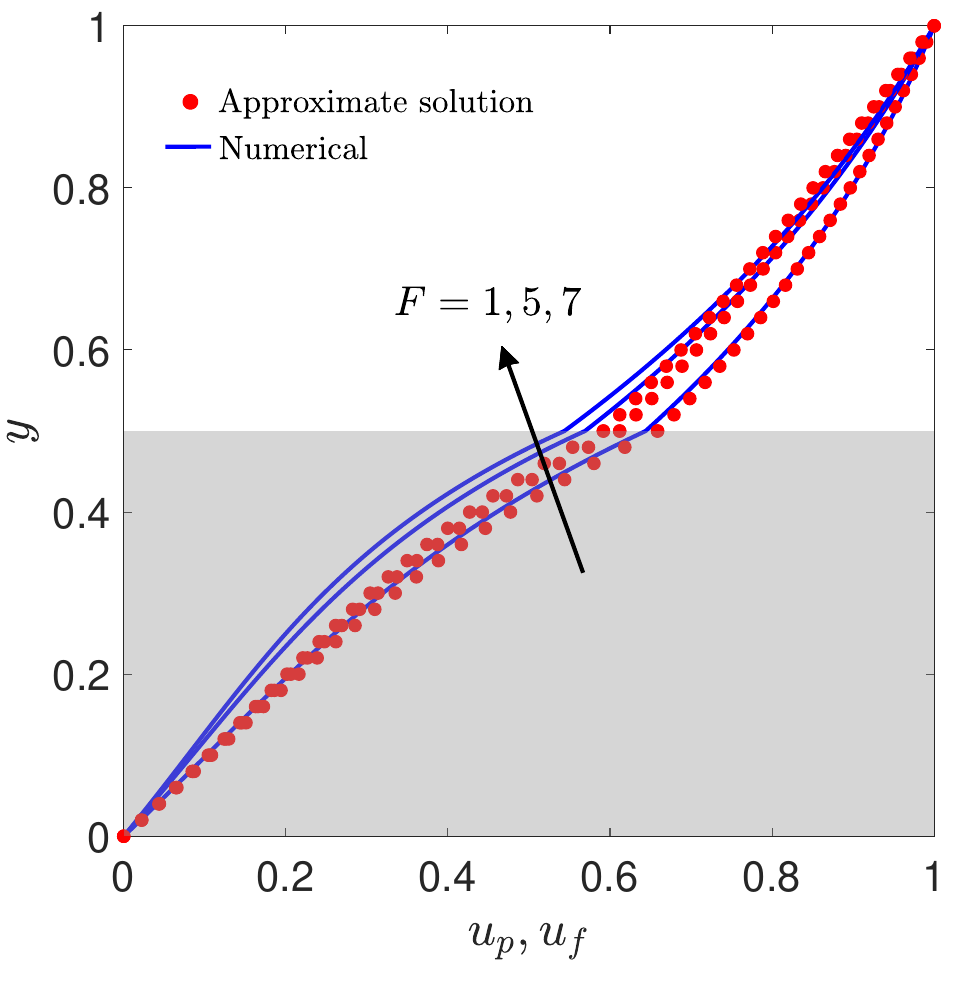}}}
    \caption{Dimensionless velocity profile (a) for different $Da$ when $\lambda=0.5$, $Q=1$, $M=1$, $\beta=0.5$, $K=1$, $F=1$, $U=1$, $f=1$, $\phi=0$; (b) for different $F$ when $\lambda=0.5$, $Q=1$, $M=1$, $\beta=0.5$, $Da=0.1$, $K=1$, $U=1$, $f=1$.}
    \label{velocity F and Da}
\end{figure}

\begin{figure}[h!]
    \centering
    {\subfigure[\label{network diagram s1c1}]{\includegraphics[height=2.5 in, width=1.5 in]{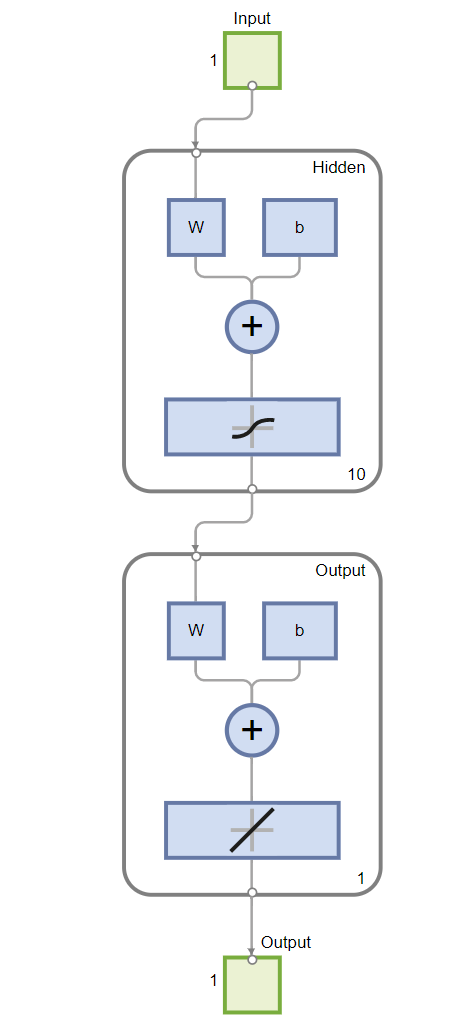}}}
    {\subfigure[\label{performance s1c1}]{\includegraphics[width=2.0 in]{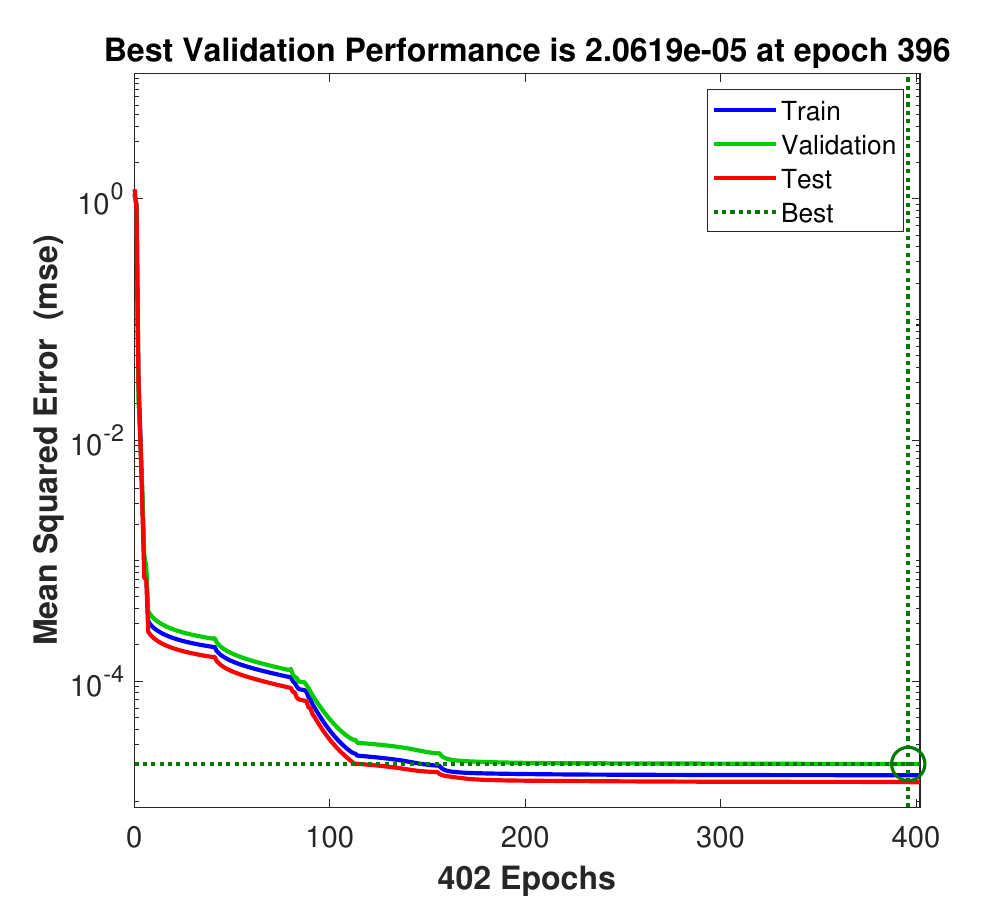}}}
    {\subfigure[\label{training state s1c1}]{\includegraphics[width=2.0 in]{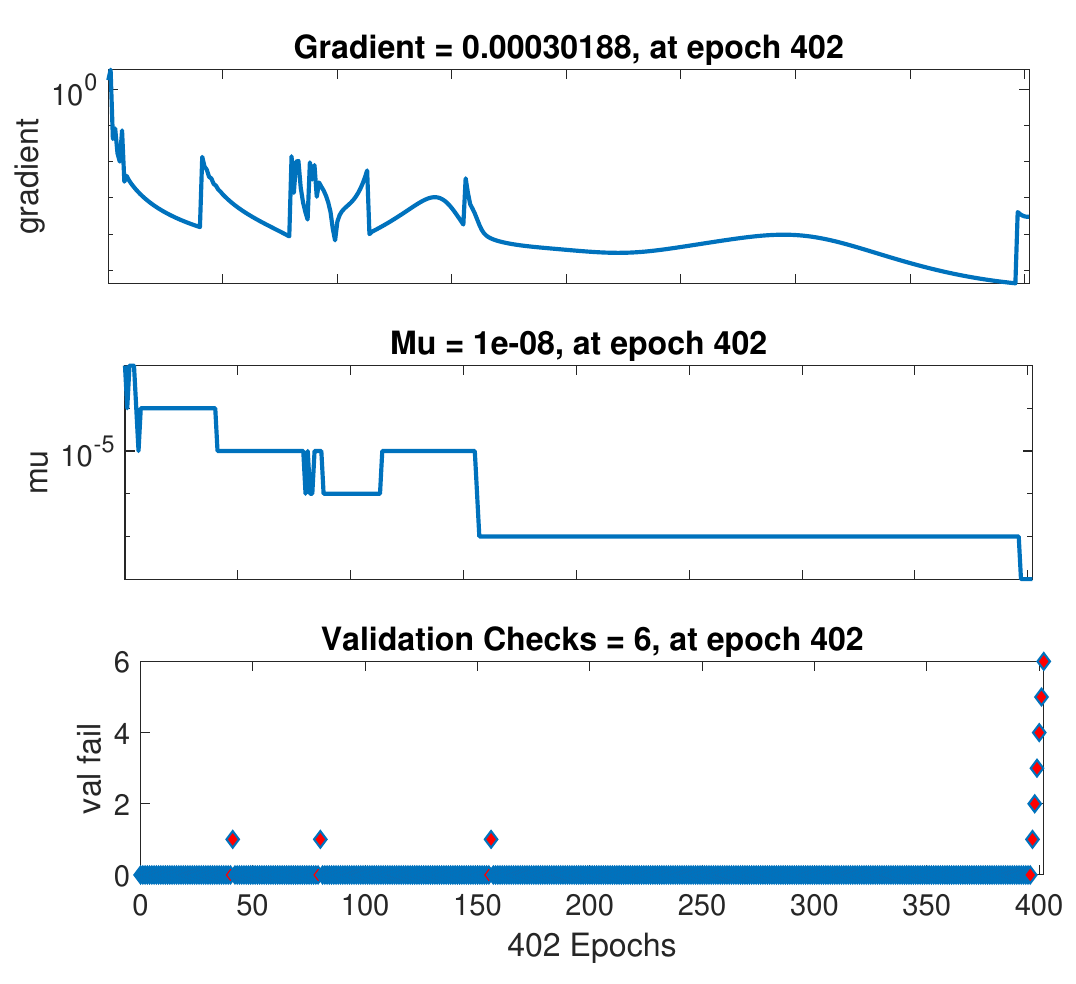}}}
    \caption{Scenario 1-Case 1.}
    \label{scenario 1 case 1}
\end{figure}

\begin{figure}[h!]
    \centering
   {\subfigure[\label{histogram s1 c1}]{\includegraphics[width=2.0 in]{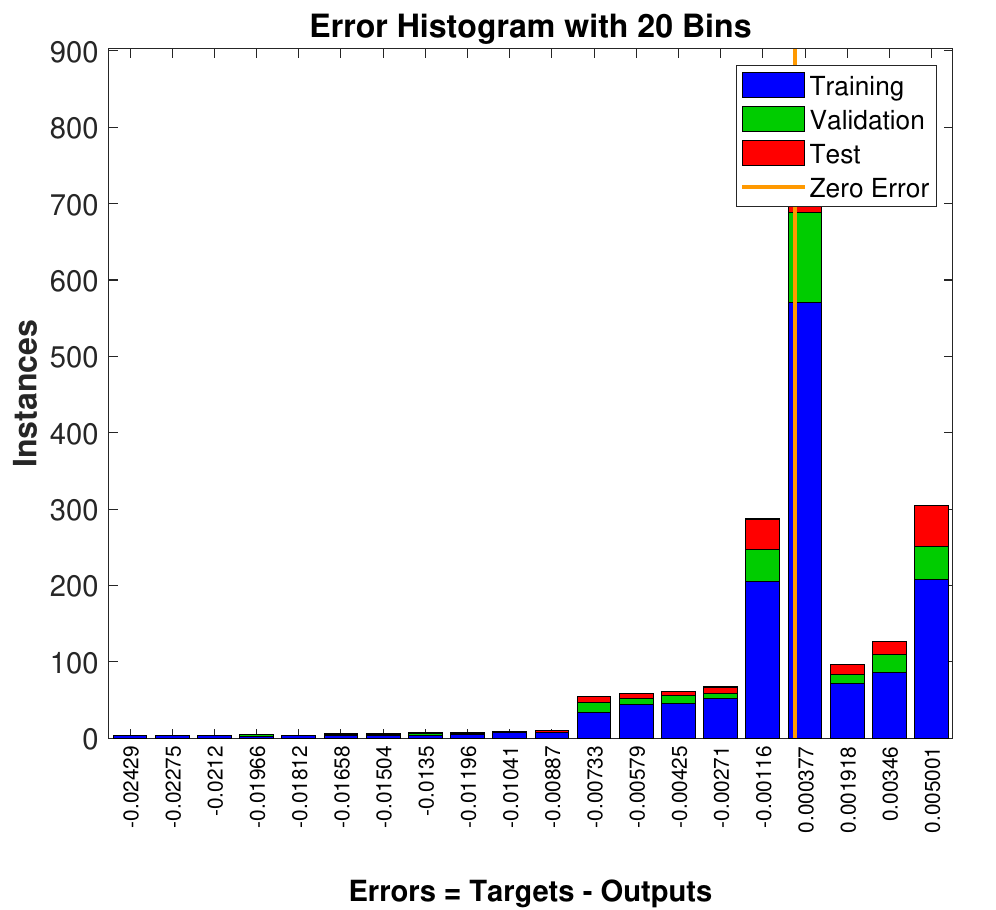}}}
    {\subfigure[\label{function fit s1 c1}]{\includegraphics[width=2.0 in]{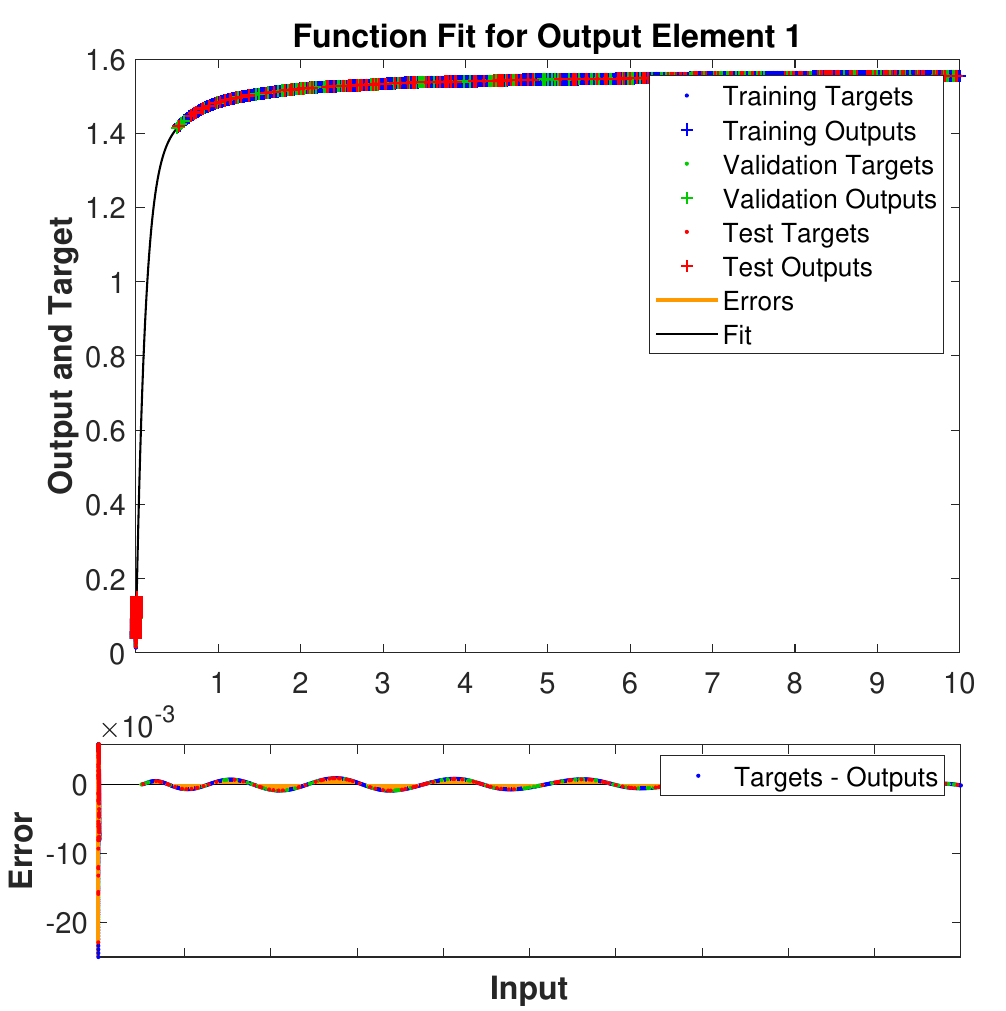}}}
    \caption{Scenario 1-Case 1.}
    \label{fig: s1 c1}
\end{figure}

\begin{figure}
    \centering
    \includegraphics[height= 3.0 in, width=3.0 in]{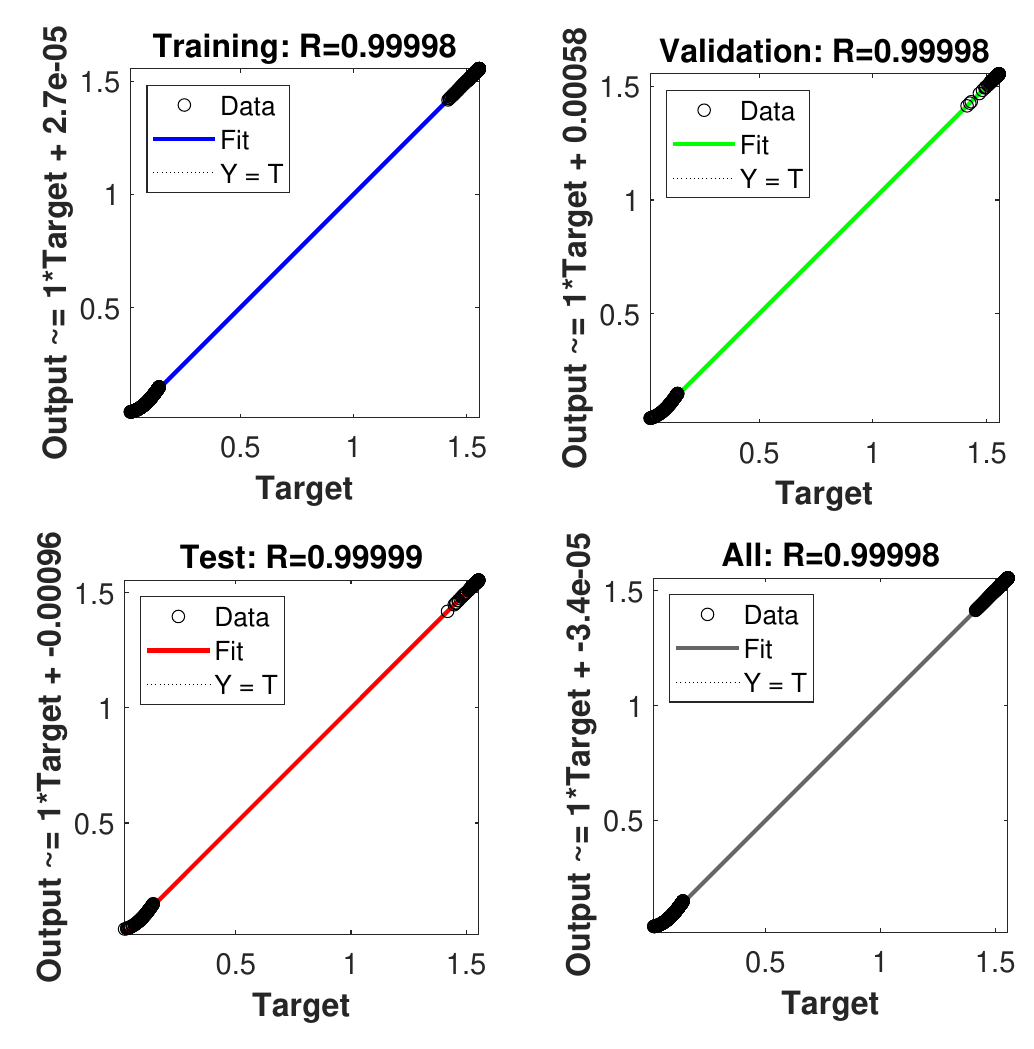}
    \caption{Scenario 1-Case 1.}
    \label{fig:regression_s1c1}
\end{figure}

Figure \ref{velocity different Da} illustrates the distribution of horizontal velocity across the channel for different Darcy numbers. Increasing the Darcy number enhances the permeability in the horizontal direction, which facilitates motion within the porous region. At lower Darcy numbers, the boundary layer is distinctly visible, and the velocity profile flattens across most of the channel's cross-section, reaching its maximum at the upper plate. The impact of the inertial coefficient on the velocity distribution is illustrated in Fig. \ref{velocity different F}. The magnitude of the velocity decreases as the inertial coefficient increases. For higher values of $F$, although the obtained approximate solution deviates slightly from the numerical solution, it still shows relatively good agreement. One may obtain better accuracy for second and higher-order iterations. 

\begin{table}[]
 \begin{center}
\begin{tabular}{  p{2.1cm} p{2.1cm} p{2.1cm} p{2.1cm}  p{2.1cm} }
\hline
Scenario & Case & $K$ & $\phi$  &$F$\\
\hline
~~ & $1$ & $0.5$ & $0$ & $1$\\
$1$ & $2$ & $1$ & $0$ & $1$\\
~~ & $3$ & $2$ & $0$ & $1$\\
\hline
~~ & $1$ & $0.5$ & $0$ & $1$\\
$2$ & $2$ & $0.5$ & $\frac{\pi}{4}$ & $1$\\
~~ & $3$ & $0.5$ & $\frac{\pi}{2}$ & $1$\\
\hline
~~ & $1$ & $1$ & $0$ & $1$\\
$3$ & $2$ & $1$ & $0$ & $5$\\
~~ & $3$ & $1$ & $0$ & $7$\\
\hline
\end{tabular}
\end{center}   
\caption{Values of the parameters in the training dataset employed for the ANN analysis}\label{Table_data s1c1}
 \end{table}

\begin{figure}[h!]
    \centering
    {\subfigure[\label{network s2c2}]{\includegraphics[height=2.5 in, width=1.5 in]{Network_Diagram}}}
   {\subfigure[\label{performance s2c2}]{\includegraphics[width=2.0 in]{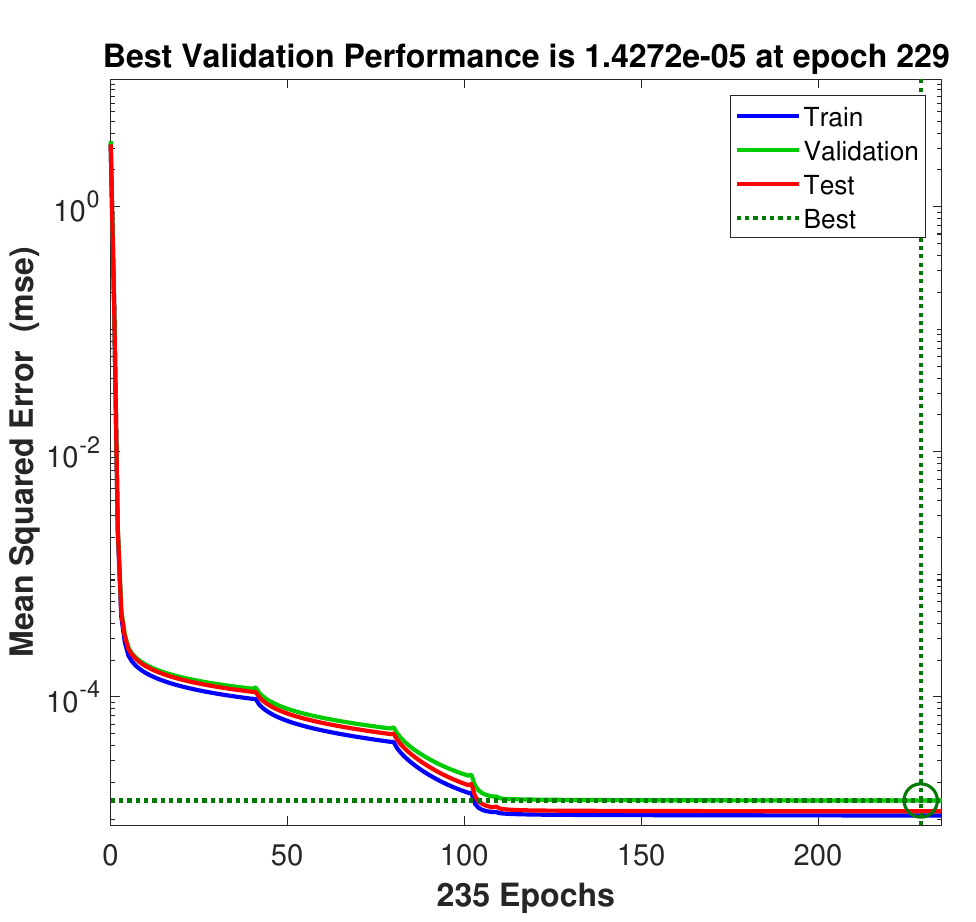}}}
    {\subfigure[\label{training state s2c2}]{\includegraphics[width=2.0 in]{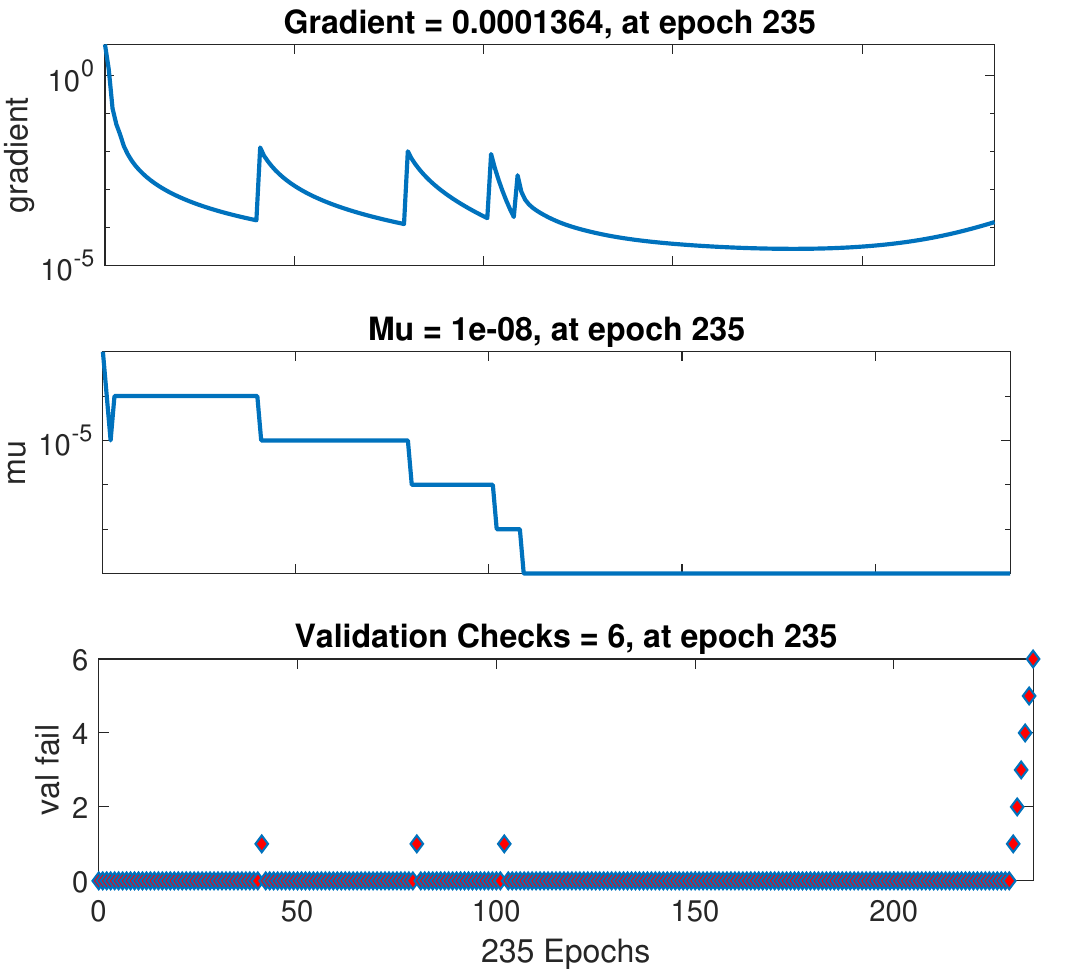}}}
    \caption{Scenario 2-Case 2.}
    \label{fig:s2c2}
\end{figure}

\begin{figure}[h!]
    \centering
   {\subfigure[\label{histogram s2c2}]{\includegraphics[width=2.0 in]{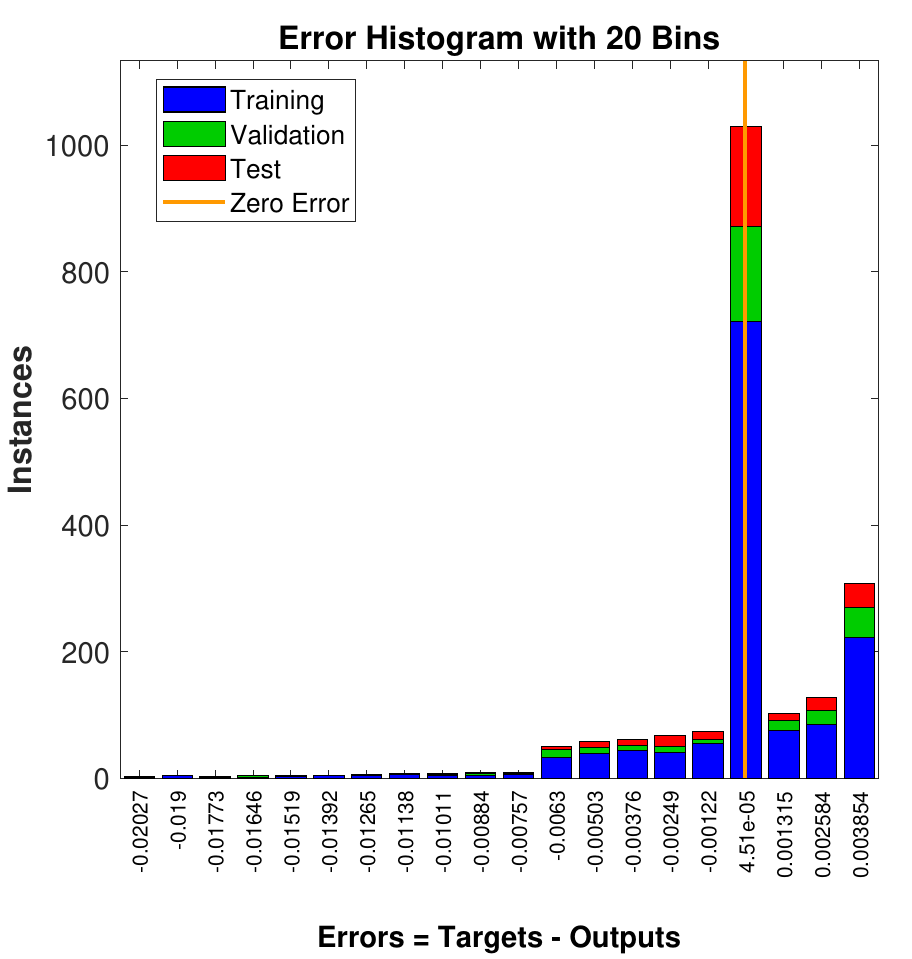}}}
    {\subfigure[\label{function fit s2c2}]{\includegraphics[width=2.0 in]{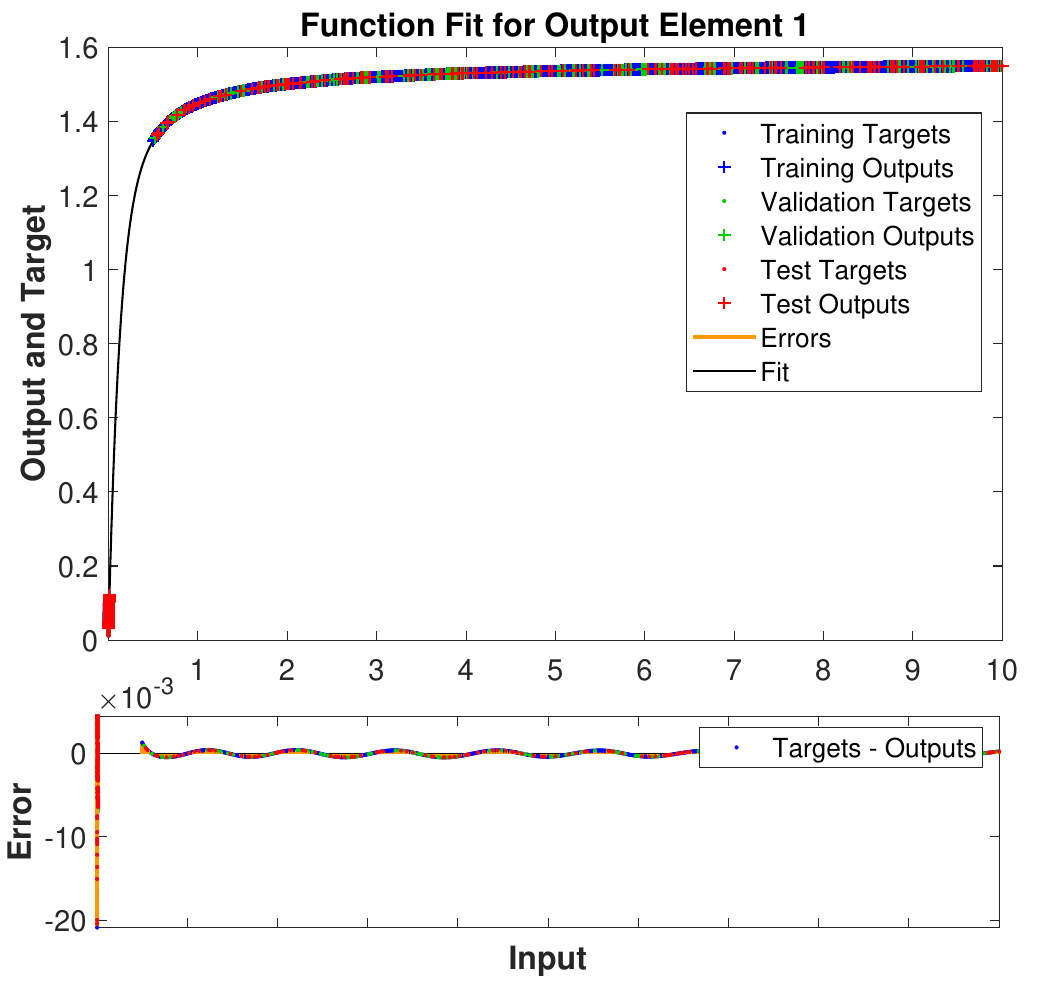}}}
    \caption{Scenario 2-Case 2.}
\end{figure}

Artificial neural networks (ANNs) have proven to be powerful tools for solving a wide range of flow problems. In many existing studies, the training datasets are generated using numerical methods. These datasets are then used to train and validate the neural network, which can subsequently predict outcomes for datasets beyond the training dataset. However, obtaining numerical solutions is often computationally expensive and challenging, particularly for complex systems. Researchers frequently rely on asymptotic or approximate solutions that are valid only within specific parameter ranges to address this. These simplified solutions allow the ANN to generalize and predict results beyond the original parameter regimes where approximate solutions are valid. In this work, we have derived the asymptotic solution for both low and high values of the Darcy number. As demonstrated in Figs. \ref{velocity low Da} and \ref{velocity high Da}, the asymptotic solution aligns well with the results for both low and high values. However, it does not provide accurate results in the moderate range of Darcy number; in such cases, one can predict the solution using a neural network beyond the specified range.

\begin{figure}
    \centering
    \includegraphics[height= 3 in, width=3 in]{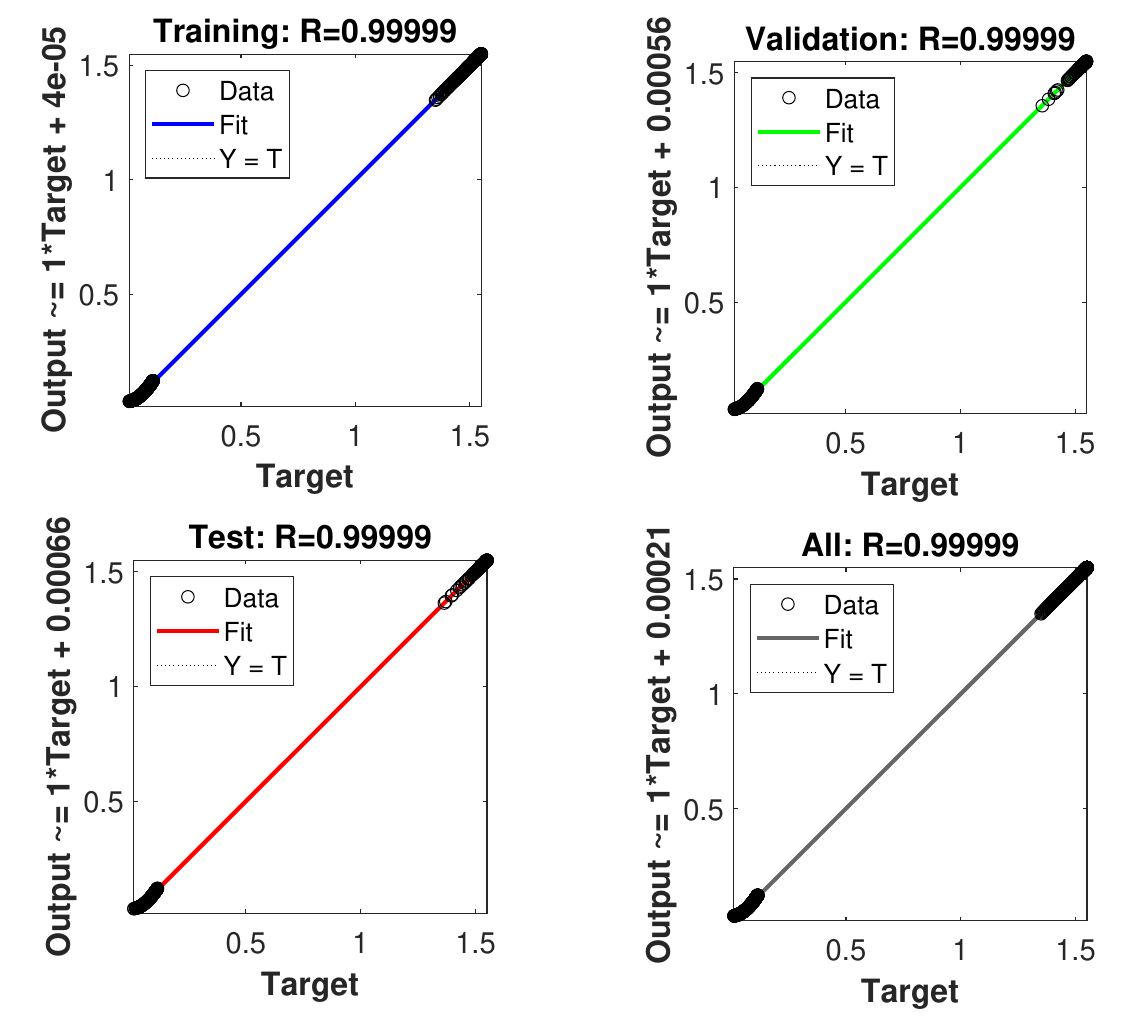}
    \caption{Scenario 2-Case 2.}
    \label{regeression s2c2}
\end{figure}

\begin{figure}[h!]
    \centering
    {\subfigure[\label{network s3c3}]{\includegraphics[height=2.5 in, width=1.5 in]{Network_Diagram}}}
   {\subfigure[\label{performance s3c3}]{\includegraphics[width=2.0 in]{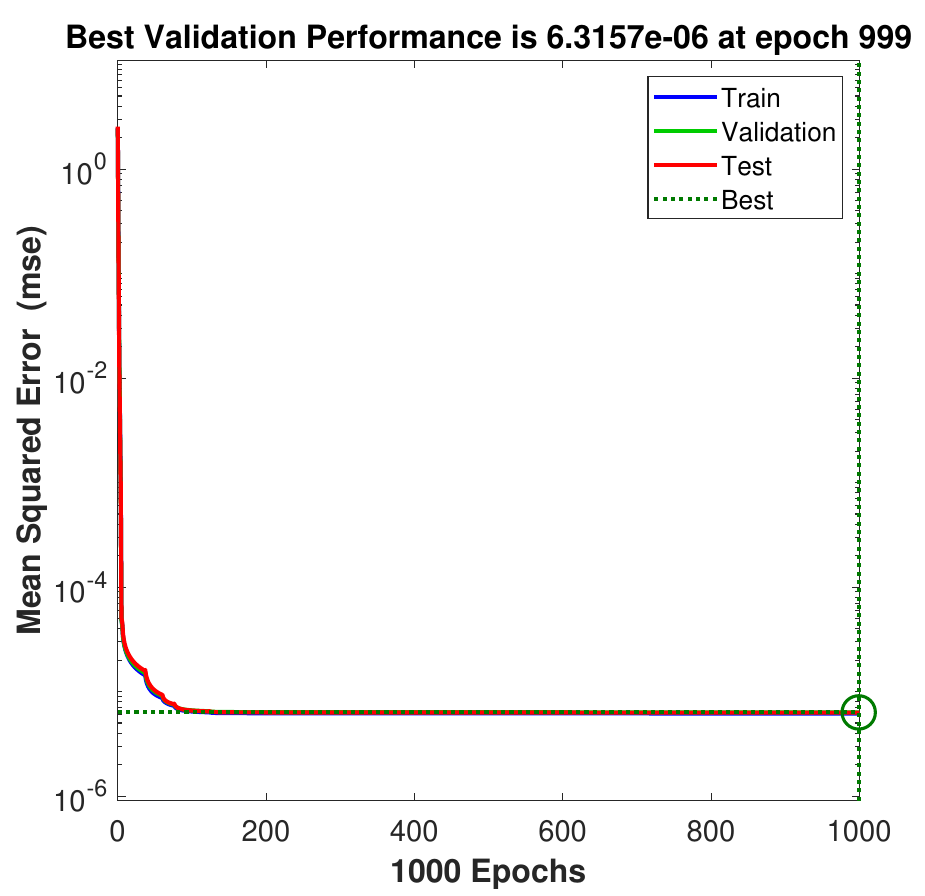}}}
    {\subfigure[\label{training state s3c3}]{\includegraphics[width=2.0 in]{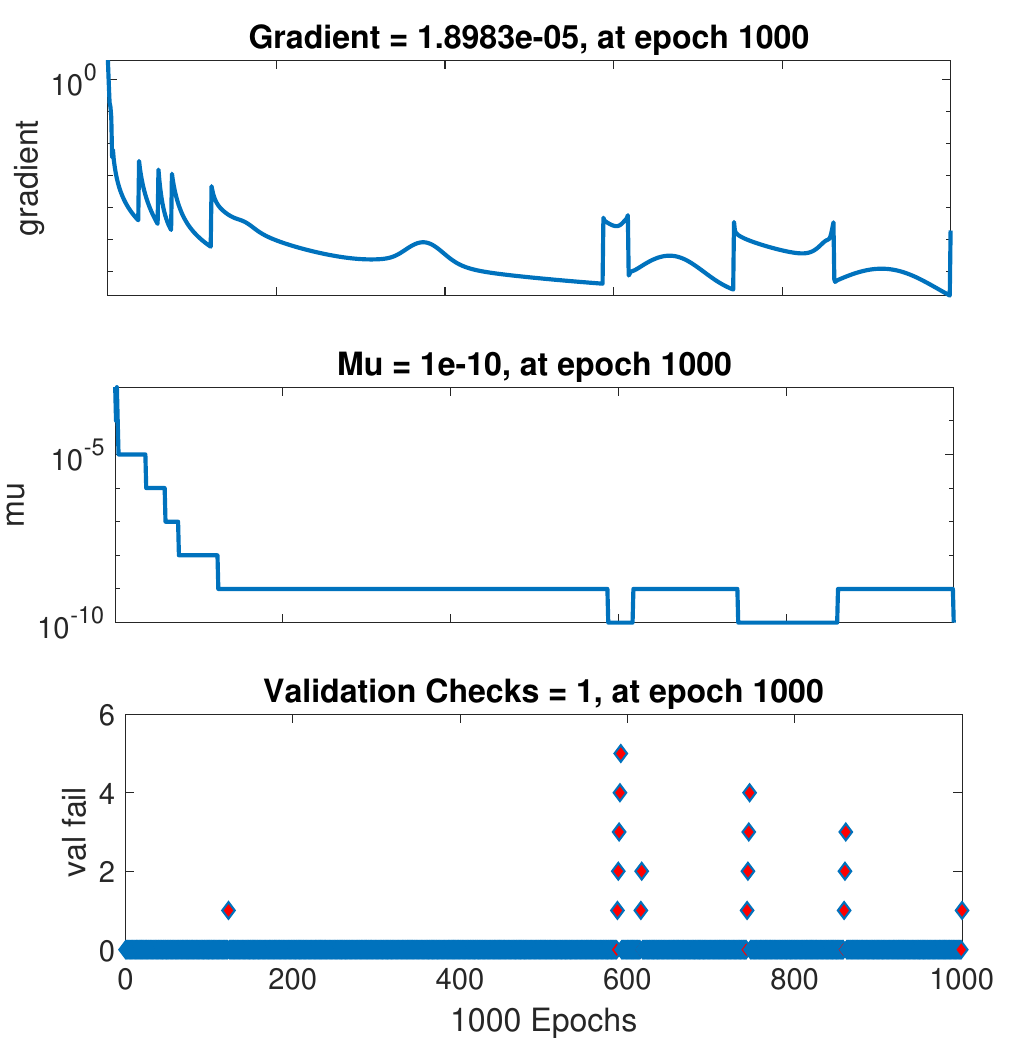}}}
    \caption{Scenario 3-Case 3.}
\end{figure}

\begin{figure}[h!]
    \centering
   {\subfigure[\label{histogram s3c3}]{\includegraphics[width=2.0 in]{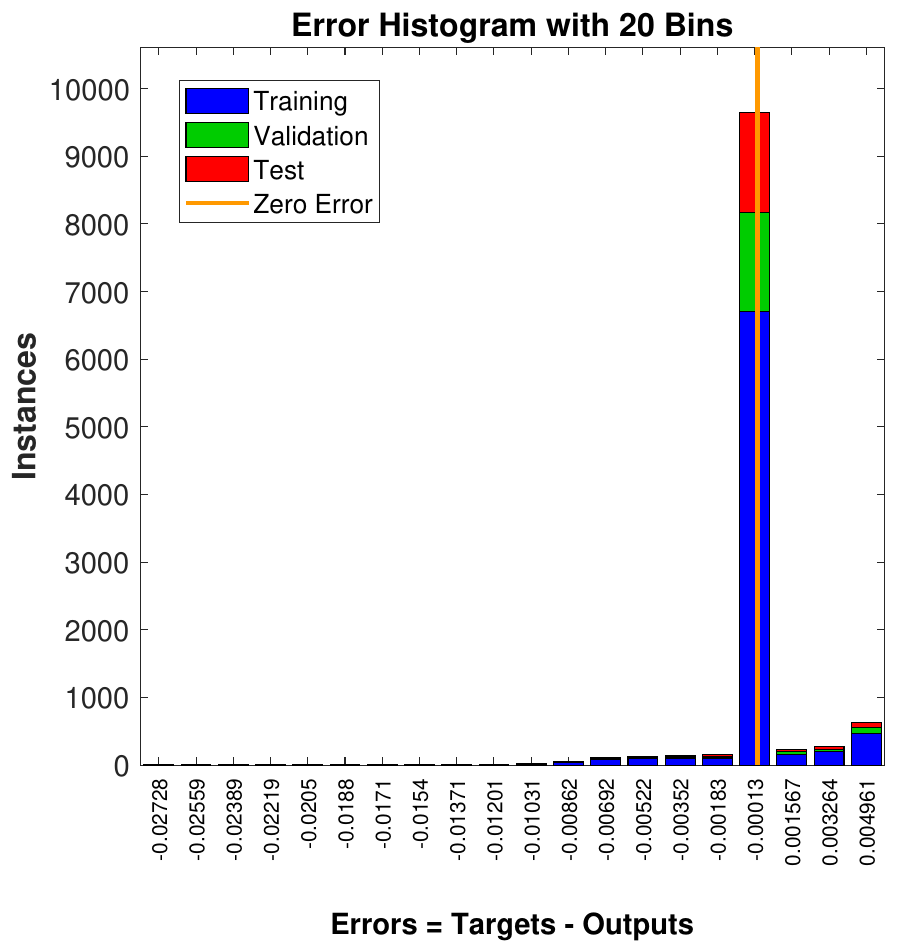}}}
    {\subfigure[\label{function fit s3c3}]{\includegraphics[width=2.0 in]{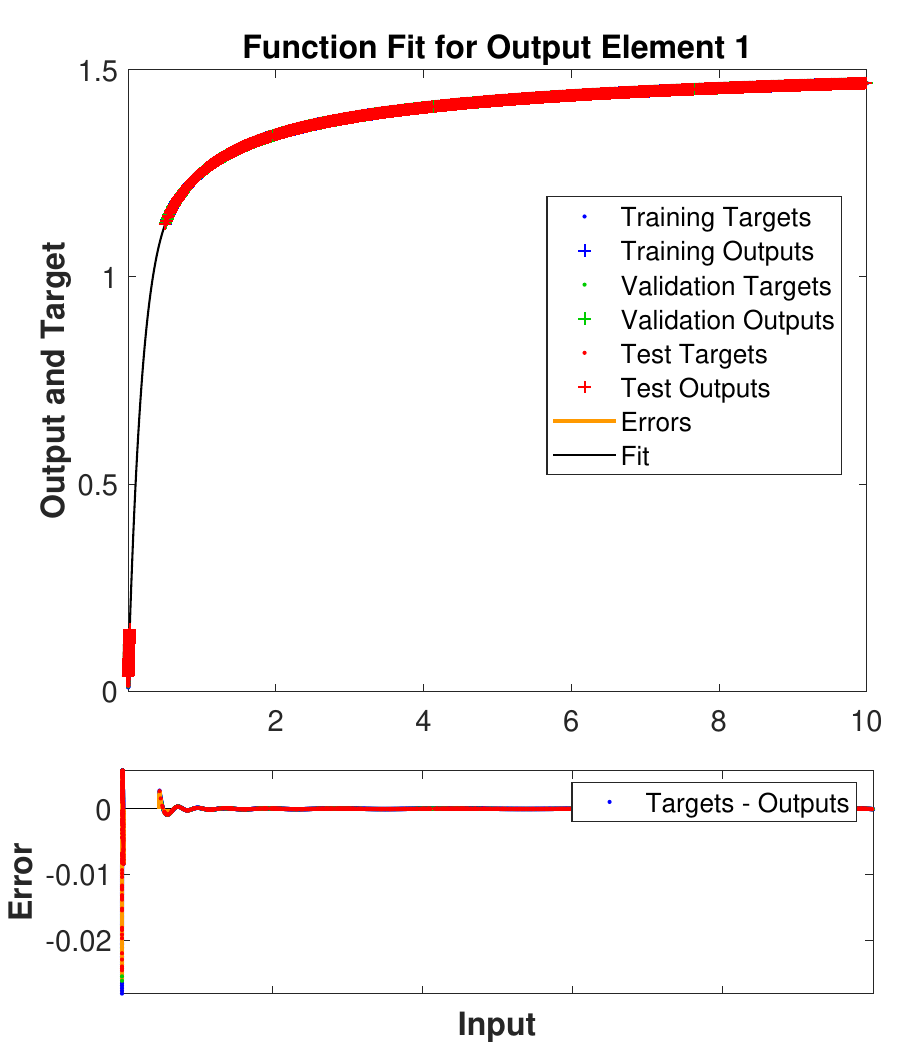}}}
    \caption{Scenario 3-Case 3.}
\end{figure}

\begin{figure}[h!]
    \centering
    \includegraphics[height= 3 in, width=3 in]{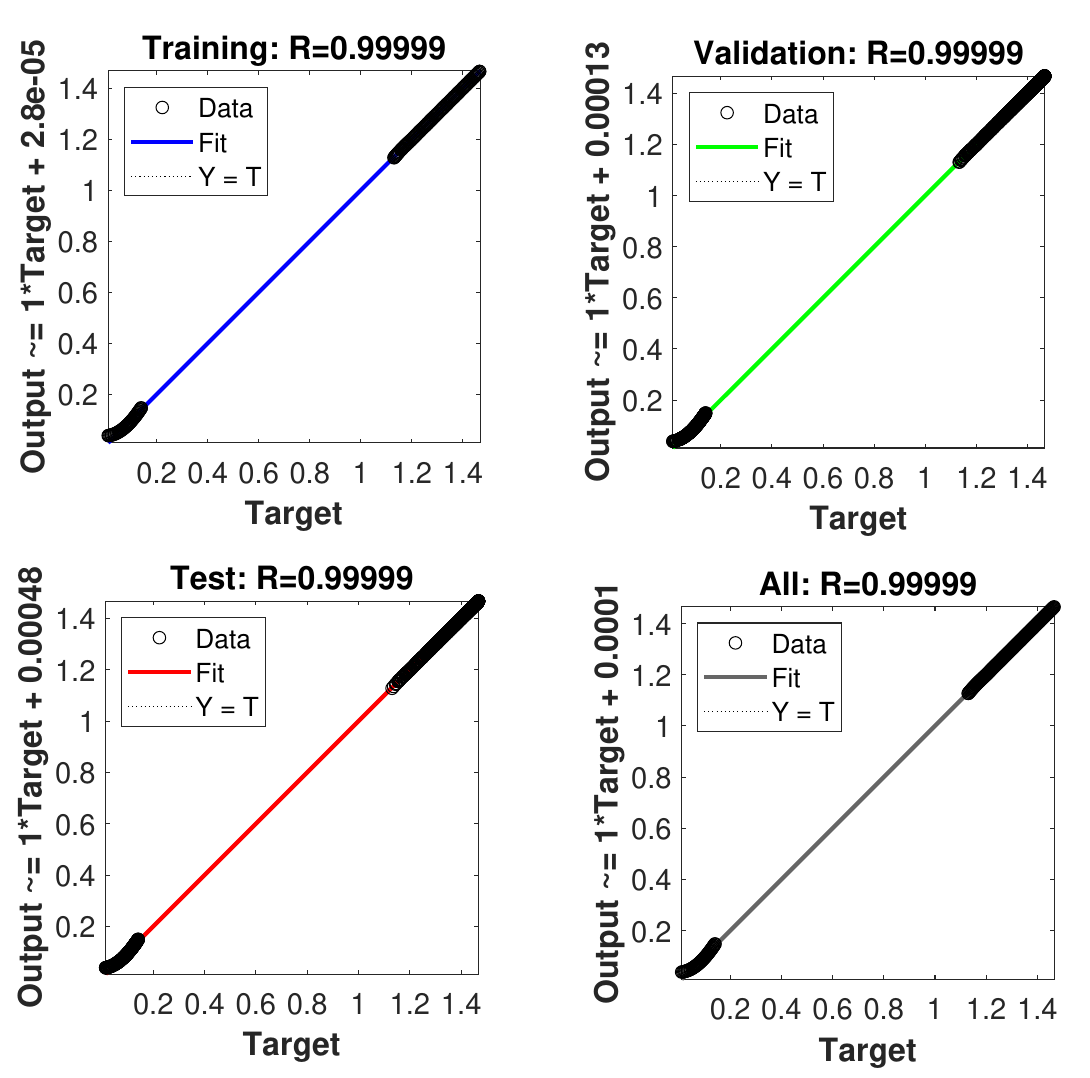}
    \caption{Scenario 3-Case 3.}
    \label{regression s3c3}
\end{figure}

\begin{figure}[h!]
    \centering
   {\subfigure[\label{shearstress diff K}]{\includegraphics[width=2.6 in]{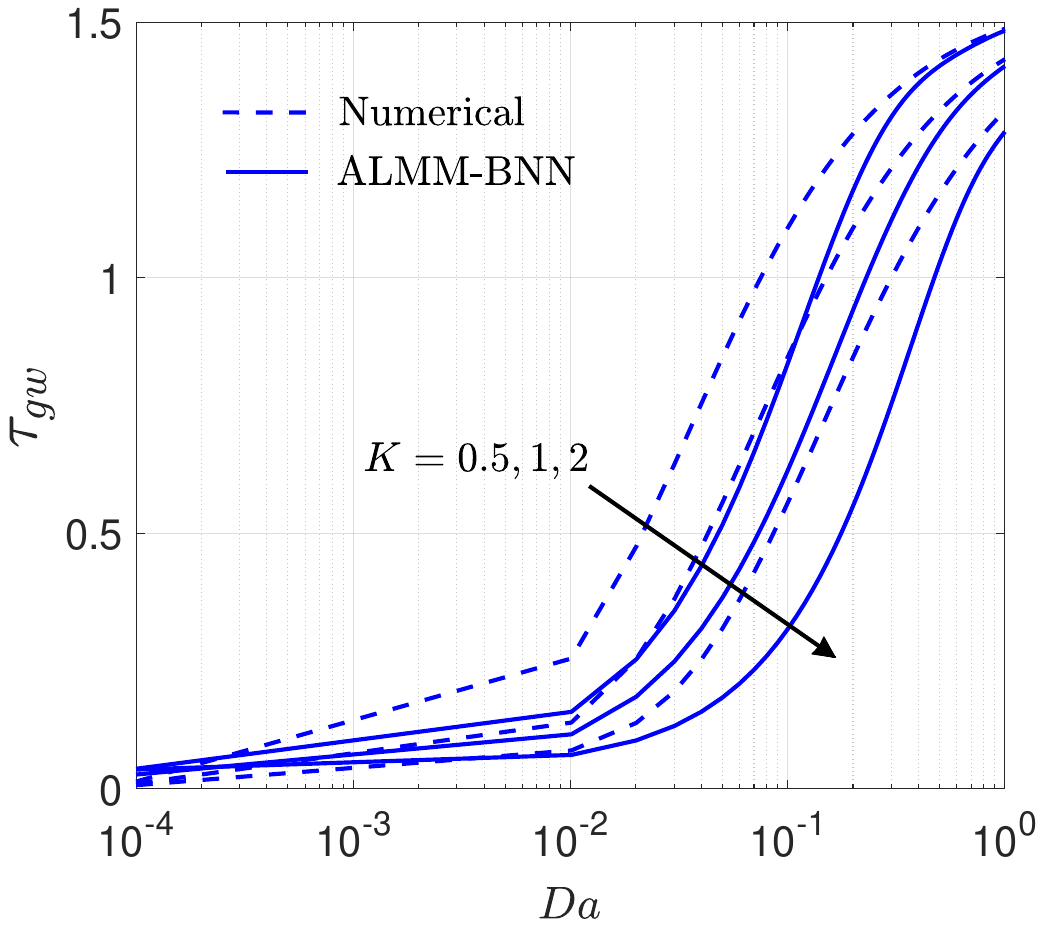}}}
    {\subfigure[\label{shearstress diff phi}]{\includegraphics[width=2.6 in]{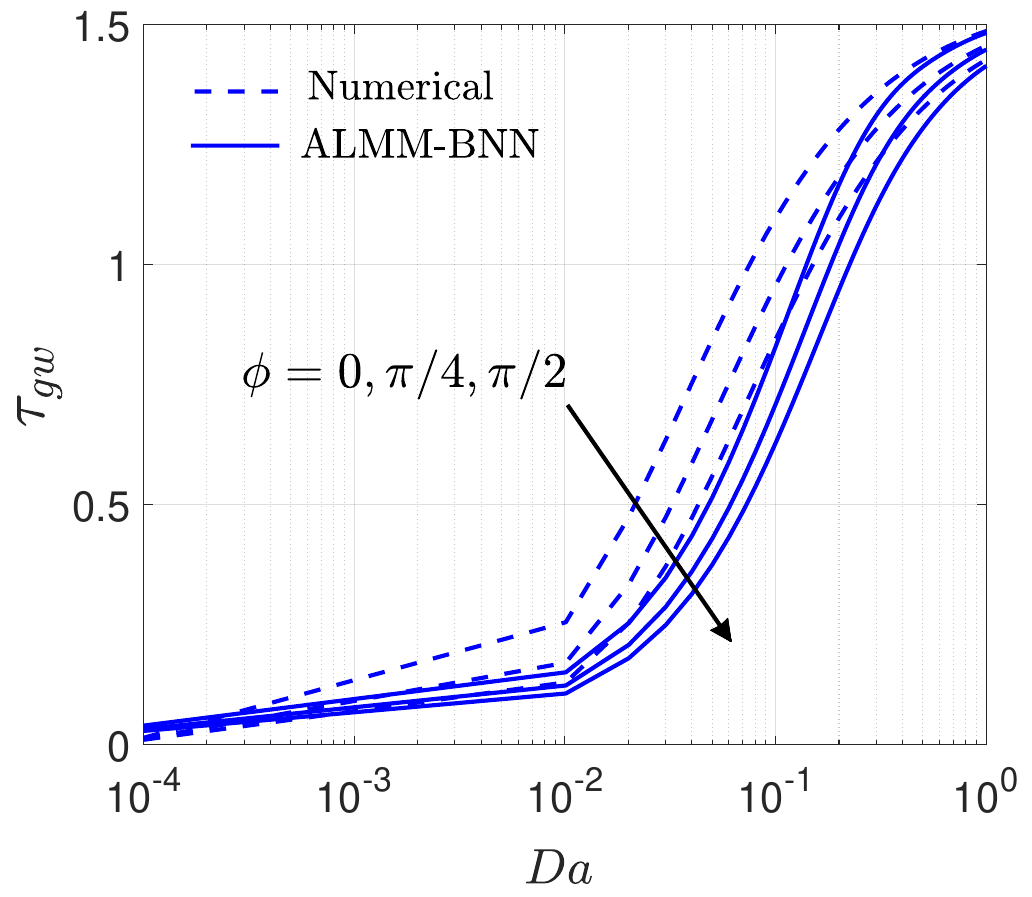}}}
    \caption{Shearstress distribution at the wall $y=0$ when $Da=0.01$, $Q=1$, $M=1$, $\beta=0$, $U=1$, $\lambda=0.5$, $F=1$ (a) for different values of $K$ when $\phi=0$; (b) for different anisotropic angle $\phi$ when $K=0.5$.}
    \label{shearstress K and phi}
\end{figure}

\begin{figure}[h!]
    \centering
   {\subfigure[\label{shearstress diff F}]{\includegraphics[width=2.6 in]{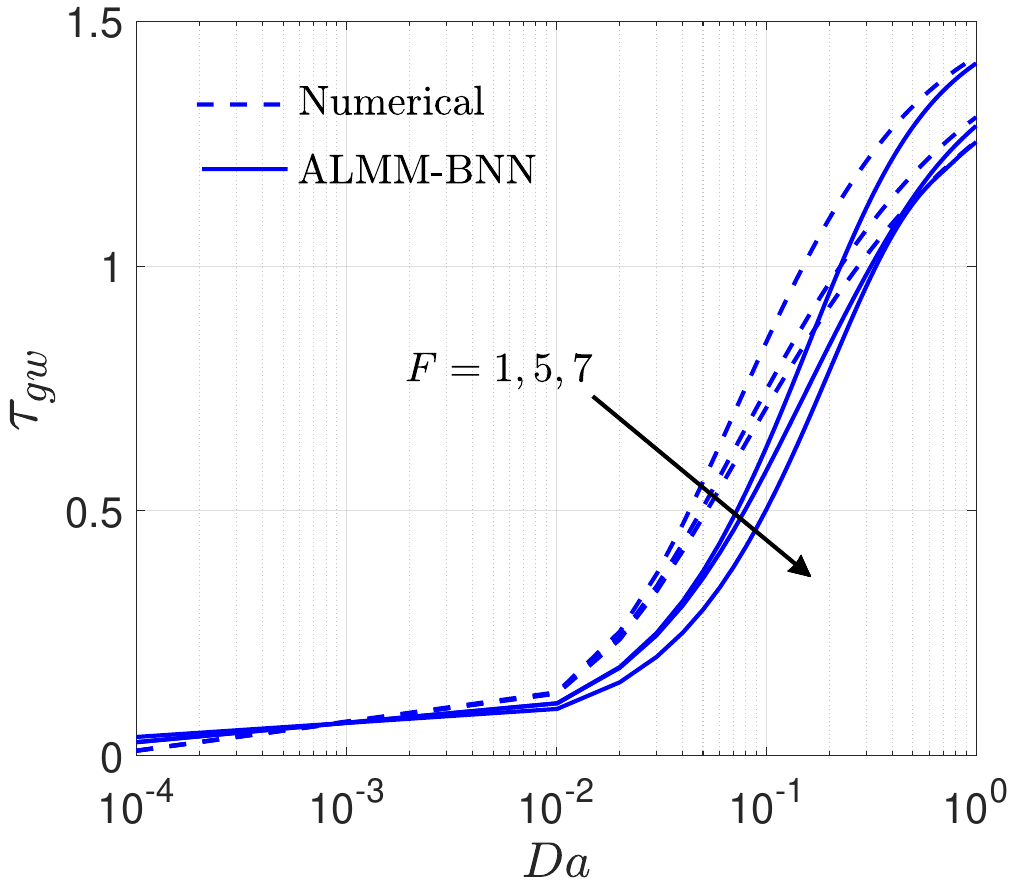}}}
    {\subfigure[\label{shearstress diff hema ratio}]{\includegraphics[width=2.6 in]{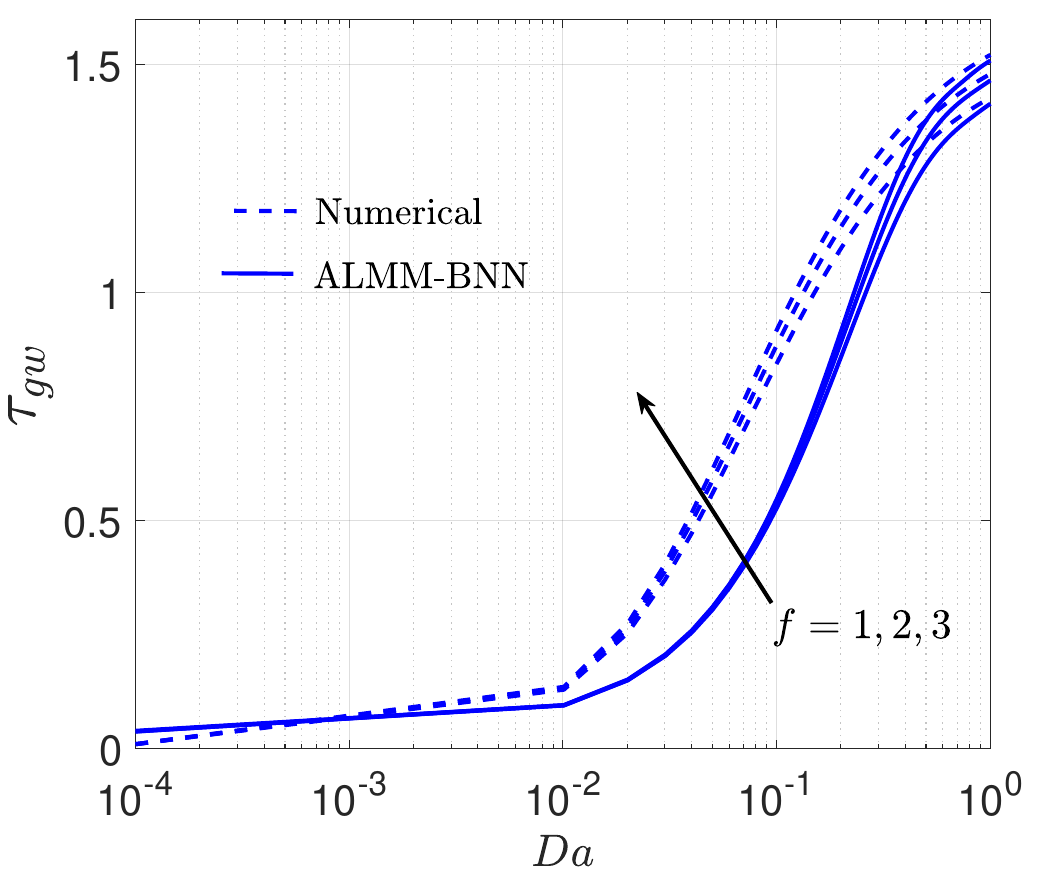}}}
    \caption{Shearstress distribution at the wall $y=0$ when $Da=0.01$, $Q=1$, $M=1$, $\beta=0$, $U=1$, $\lambda=0.5$, $K=1$, $\phi=0$ (a) for different values of $F$ when $f=1$; (b) for different $f$ when $F=1$.}
    \label{shearstress F and f}
\end{figure}

We leverage the neural network fitting toolbox `nftool', a MATLAB function that opens neural net fitting app, to predict the outcomes from the dataset based on a backpropagation neural network (BNN). The tool employs the artificial Levenberg-Marquardt method, an advanced optimization algorithm that combines the rapid convergence of the Gauss-Newton method and the stability of the gradient descent method. This hybrid technique is particularly effective, without having to compute the Hessian matrix, for minimizing the mean squared error (MSE) in a nonlinear system, making it an excellent choice for complex pattern recognition and regression tasks. This tool optimizes the network by using a three-phase structured approach: training, testing, and validation with the independent data sets. The training process requires more data compared to validation and testing. In the training process, a large amount of data is used to train the neural network, which allows the Levenberg-Marquardt method to adjust the network's weights and biases in such a way that the error between the predicted output and the actual output is minimized.

The validation phase helps to monitor the network's performance over different model types and hyperparameter choices. This process helps in determining the appropriate stopping criteria by checking the validation error and hence prevents overfitting. It also indicates whether the network is learning noise rather than a generalized pattern. During the testing phase, our model's unbiased accuracy is estimated, and the model's performance is evaluated using data it has not encountered during the training and validation. The reference dataset is obtained based on the asymptotic solution for the parameter combination given in the Table.~\ref{Table_data s1c1} for all the subsequent cases.

Figure \ref{network diagram s1c1} illustrates the architecture of a three-layer neural network scheme, which suitable for a regression task, consisting of one input layer, ten hidden layers, and one output layer. The hidden layer is functioned with hidden neurons as the sigmoid transfer function \(\frac{1}{1+e^{-x}}\), and there is a linear transformation function in the output layer. In the current ALMM-BNN-based model, $70\%$ dataset is used for training, $15\%$ for testing, and $15\%$ for validation. The reference dataset for this is obtained from Scenario 1-Case 1.
Figure~\ref{performance s1c1} exhibits the graphical representation of the mean squared error (MSE). The optimal performance is observed at epoch $396$, where MSE reaches approximately $2.0619 \times 10^{-5}.$ Figure~\ref{training state s1c1} depicts the state transition plot showing how the model's state, namely, gradient, $\textrm{Mu}$, and validation check varies with each iteration. At epoch $402$, we see that the gradient is $0.00030188$ and $\textrm{Mu}$ is $10^{-8}$. Figure \ref{histogram s1 c1} shows that the machine learning model is highly accurate, with most errors concentrated around zero, indicating that most predictions are close to the target value. These findings are supported by the function fit plot in Fig.~\ref{function fit s1 c1}. The regression value $R$ is a statistical measure indicating the relationship between the predictions (output) and the responses (target) for the training, validation, and test data sets. An $R$ value close to unity signifies that all the variance in the data is well explained and there is a strong positive correlation meaning the model's prediction aligns exactly with the true values. The regression plots in Fig.~\ref{fig:regression_s1c1} exhibit correlation value of $R\approx1$ between the output and target values, indicating a strong performance of the ALMM-BNN model across the training, validation, and test datasets.

We have obtained the reference dataset for the parameter values outlined in Scenario 2-Case 2, as shown in Figures \ref{network s2c2} -\ref{regeression s2c2}, as well as for Scenario 3-Case 3, presented in Figures \ref{network s3c3} -\ref{regression s3c3}. Additionally, we have conducted other case studies listed in Table \ref{Table_data s1c1}; however, those results are not included here due to space constraints.


To evaluate the accuracy and advantages of the ALMM-BNN method, we have plotted the shear stress distribution at the lower wall, denoted as \(\tau_{gw} = \frac{du_{p}}{dy}|_{y=0}\), against the Darcy number in Fig.~\ref{shearstress K and phi}. Within the range \(10^{-4} \leq Da \leq 10^{-2}\), we obtained the shear stress distribution using the matched asymptotic solution described in Section \ref{solution small Darcy}. For \(0.5 \leq Da \leq 1\), we calculate the shear stress using a regular perturbation solution outlined in Section \ref{large Darcy}. In the moderate range, specifically when \(10^{-2} \leq Da \leq 0.5\), we predict the solution using the ALMM-BNN paradigm. Fig. \ref{shearstress diff K} depicts the shear stress distribution for different anisotropic permeability ratios, $K$. We see that the wall shear stress increases as the permeability ratio $K$ decreases. This happens because for a fixed value of $Da$ (i.e., $K_{1}$), a reduction in the anisotropic ratio results in decreased permeability, leading to lower velocity and consequently lower shear stress. The effect of the orientation angle $\phi$ on the distribution of wall shear stress is shown in Fig. \ref{shearstress diff phi}. As the orientation angle increases, there is a noticeable reduction in shear stress, which can be attributed to the decrease in flow velocity. Figure \ref{shearstress diff F} illustrates the shear stress distribution for different inertial coefficients $F$.  We see that a larger $F$ hinders the flow, causing it to move more slowly, which results in a reduction of shear stress. This effect is more noticeable at higher values of $Da$, where the influence of flow inertia becomes significant. We see that the predicted solution shows a qualitative alignment with the numerical solution. Figure \ref{shearstress diff hema ratio} illustrates the shear stress variation at the bottom wall as a function of the Darcy number. As the parameter $f$ increases, the wall shear stress also increases, with the effect being more pronounced at higher Darcy numbers. This finding is particularly relevant for understanding shear stress distribution during blood flow in arteries. These results indicate that a higher hematocrit ratio increases shear stress exerted on the endothelial cell layer. In all these results, the predicted solution obtained using the ALMM-BNN framework demonstrates qualitatively good agreement with the numerical results. It's interesting to see that, while the predicted values do not perfectly coincide with the numerical solutions, they successfully capture the underlying trends as the flow parameters vary. This consistency highlights a key advantage of the ALMM-BNN approach: it provides reliable approximations even when numerical solutions are unavailable or computationally expensive.

\begin{figure}
    \centering
    \includegraphics[width=3 in]{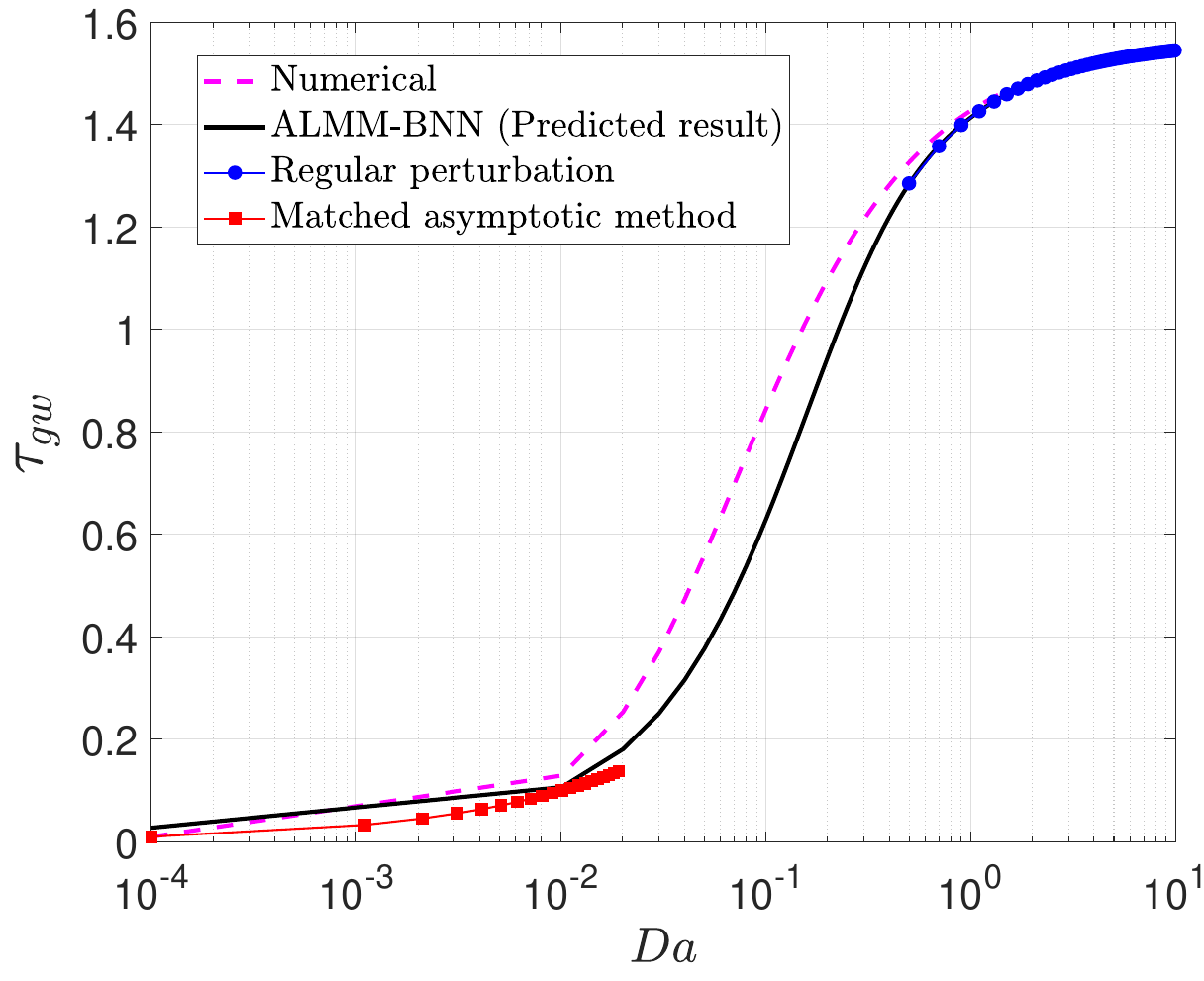}
    \caption{Shear stress distribution at the wall $y=0$ when $Da=0.01$, $Q=1$, $M=1$, $\beta=0$, $U=1$, $\lambda=0.5$, $K=1$, $F=1$.}
    \label{shearstress all compare}
\end{figure}

To get a complete picture of the predicted, numerical, and perturbation solutions, we plot the shear stress distribution in Fig. \ref{shearstress all compare}. For both high and low values of the Darcy number, asymptotic solutions obtained through regular perturbation and matched asymptotic expansion show good agreement with the numerical results. However, these traditional analytical methods are not valid in the intermediate range, e.g., $10^{-2}\leq Da \leq 0.5$. In this regime, the ALMM-BNN paradigm offers a valuable alternative. While it may not yield exact solutions, it effectively captures the overall trend and provides qualitatively good agreement with the numerical results. This highlights the potential of the ALMM-BNN paradigm as a robust predictive tool in challenging parameter ranges where numerical solutions are either difficult to obtain or computationally expensive.

\section{Conclusions}
In this study, we investigate the Couette-Poiseuille flow of a fluid with variable viscosity in a channel that is partially filled with an anisotropic porous medium. We have used the Navier-Stokes equation to model the free flow region and the Brinkman-Forchheimer extended Darcy's law that governs the flow inside the porous media, which accounts for the inertia. The resulting liquid-porous coupled system exhibits nonlinearity and depth-dependent viscosity, presenting significant analytical and numerical challenges. We get an asymptotic solution for the nonlinear coupled Navier-Stokes and Brinkman-Forchheimer extended Darcy's equation by considering the Darcy number as a perturbed parameter. For a large Darcy number, we obtain the corresponding solution using a regular perturbation expansion. In the case of a low Darcy number, the analogous problem of interest is a singular perturbation problem, which is dealt with using the Prandtl matching principle. These asymptotic methods aligned well with the numerical results at high and low parameter limits but proved inadequate for the intermediate range. To bridge the gap, we have used an artificial Levenberg-Marquardt method with a backpropagated neural network (ALMM-BNN), which may not yield an accurate solution; however, it effectively captures the flow's qualitative behavior and provides a reliable prediction where the traditional asymptotic method fails. In addition to the asymptotic solution, we presented an approximate solution based on an iterative method that aligns with the numerical solution across a wider range of parameter values. A detailed synopsis is provided on the effect of anisotropic permeability within a porous medium that includes a non-linear Forchheimer inertial term. This model gives valuable insight into understanding the shear stress distribution in high-speed blood flow in arteries where the inertial term is significant.







\providecommand{\noopsort}[1]{}\providecommand{\singleletter}[1]{#1}%
%



\end{document}